\documentclass[a4paper,UKenglish,cleveref, autoref, thm-restate]{lipics-v2019}
\pdfoutput=1
\usepackage{macros}
\usepackage{diagrams-ogog}
\nolinenumbers
\title{Compositional modelling of network games}

\author{Elena Di Lavore}{Department of Software Science, Tallinn University of Technology, Estonia}{elena.di@taltech.ee}{0000-0002-7783-5079}{}
\author{Jules Hedges}{Department of Computer and Information Sciences, University of Strathclyde, Scotland}{jules.hedges@strath.ac.uk}{[orcid]}{}
\author{Pawe\l{} \ Soboci\'nski}{Department of Software Science, Tallinn University of Technology, Estonia}{pawel.sobocinski@taltech.ee}{[orcid]}{}
\funding{\emph{Elena Di Lavore} and \emph{Pawel Sobocinski} are supported by the ESF funded Estonian IT Academy research measure (project 2014-2020.4.05.19-0001)}
\authorrunning{E.\ Di Lavore, J.\ Hedges and P.\ Soboci\'{n}ski}
\Copyright{Elena Di Lavore, Jules Hedges and Pawe{\l} Soboci\'{n}ski}
\EventEditors{Christel Baier and Jean Goubault-Larrecq}
\EventNoEds{2}
\EventLongTitle{29th EACSL Annual Conference on Computer Science Logic (CSL 2021)}
\EventShortTitle{CSL 2021}
\EventAcronym{CSL}
\EventYear{2021}
\EventDate{January 25--28, 2021}
\EventLocation{Ljubljana, Slovenia (Virtual Conference)}
\EventLogo{}
\SeriesVolume{183}
\ArticleNo{20}
\begin{document}
\maketitle
\begin{abstract}
%\abstract{
  The analysis of games played on graph-like
  structures is of increasing importance due to the prevalence of
  social networks, both virtual and physical, in our daily life.
  As well as being relevant in computer science, mathematical
  analysis and computer simulations of such distributed games are
  vital methodologies in economics, politics and epidemiology,
  amongst other fields. Our contribution is to give compositional
  semantics of a family of such games as a well-behaved mapping,
  a strict monoidal functor, from a category of open graphs
  (syntax) to a category of open games (semantics). As well as
  introducing the theoretical framework, we identify some
  applications of compositionality.
%}
\end{abstract}

\section{Introduction}

%compositionality grand intro
Compositionality concerns finding homomorphic mappings
\begin{equation}\label{eq:compositionality}
\mathbf{Syntax} \rightarrow \mathbf{Semantics}.
\end{equation}
This important concept originated in formal logic~\cite{Tarski1933,Tarski1957}, and is at the centre of formal semantics of programming languages~\cite{Winskel1993}. In recent years, there have been several \emph{2-dimensional} examples~\cite{Bruni2011,Bonchi2015a,Baez2015a}, where both $\mathbf{Syntax}$ and $\mathbf{Semantics}$ are symmetric monoidal categories. Usually $\mathbf{Syntax}$ is freely generated from a (monoidal) signature, possibly modulo equations. This opens up the possibility of recursive definitions and proofs by structural induction, familiar from our experience with ordinary, 1-dimensional syntax.

%network games as a surprising instance
\medskip
In this paper, we consider an instance of~\eqref{eq:compositionality} that is---at first sight---quite different from the usual concerns of programming and logic: network games~\cite{bramoulle_handbook_economics_network}, also known as graphical games. Network games involve agents that play concurrently, and share information based on an underlying, ambient network topology.
Indeed, the utility of each player typically depends on the structure of the network. An interesting
%\todo{not obvious to reviewer 2}
application is social networks~\cite{spread_of_happiness}, but they also feature in economics~\cite{galeotti_searching_pricing}, politics~\cite{unemployment_network_games} and epidemiology~\cite{epidemy_network_games}, amongst other fields.
(These games should not be confused with classes of dynamic games played on graphs, such as parity games and pursuit games, which are not within the scope of this paper.)

%Social networks influence many aspects of our lives. Often our choices on which career to take up, whether to buy a product or whether to attend a party are influenced by our professional and social connections.

%As an example, we can consider a situation in which players obtain their maximum utility when they take the same choice that the majority of their neighbours take. It can be a simple model for situations in which players are people trying to be recognised as part of a community by adapting to a common behaviour\doubt{cite something with empirical evidence? find a more well-known example?}. This model is called a \emph{majority game} and will be analysed through the paper.

% network games as games on graphs
\smallskip
In formal accounts of network games, graphs represent network topologies. Players are identified with graph vertices, and their utility is influenced only by their choices and those of their immediate neighbours. Network games are thus ``games on graphs''. An example is the \emph{majority game}: players ``win'' when they agree with the majority of their neighbours.

% network games as an example of compositionality
\medskip
In what way do such games fit into the conceptual framework of~\eqref{eq:compositionality}?
%In this paper we propose a categorical approach to games on graphs.
Our main contribution is the framing of certain network games as monoidal functors from a suitable category of \emph{open graphs}
$\Graph$~\cite{chantawibul_sobocinski_towards_compositional_graph_theory}, our $\mathbf{Syntax}$, to the category of open games $\Game$~\cite{hedges_etal_compositional_game_theory}, our $\mathbf{Semantics}$. Given a network game $\N$ (e.g.\ the majority game), such games are functors
\begin{equation}\label{eq:Fn}
F_\N \colon \Graph \rightarrow \Game
\end{equation}
that, for any \emph{closed} graph $\Gamma\in\Graph$, yield the game $F_\N(\Gamma)$, which is the game $\N$ played on $\Gamma$.
However, compositionality means that such games are actually ``glued together'' from simpler, \emph{open} games. In fact, $F_\N$ maps each vertex of $\Gamma$ to an open game called the \emph{utility-maximising player}, and the connectivity of $\Gamma$ is mapped, following the rules of $\N$, to structure in $\Game$.

% advantages
Our contribution thus makes the intuitively obvious idea that the data of network games is dependent on their network topology precise. Concrete  descriptions of network games, given a fixed topology, are often quite involved: our approach means that they can be derived in a principled way from basic building blocks. In some cases, the compositional description can also help in the mathematical analysis of games.
For example, in the case of the majority game, the right decomposition of a network topology $\Gamma$ as an expression in $\Graph$ can yield a recipe for the Nash equilibrium of $F_\N(\Gamma)$ in $\Game$ in terms of the equilibria of the open games obtained via $F_\N$ from the open graphs  in the decomposition. As it happens when solving optimization problems, a compositional analysis of the equilibria is possible only when the game has optimal substructure, which is the case for the majority game (but is not the case in general). Nevertheless, compositional modelling is valuable for the understanding of the structure of the system. It allows, for example, to modify a part of a system while keeping the analysis done for the rest of the system, as we show in Example~\ref{ex_f:best-shot_public_goods}.
%\todo{expanded about compositional modelling rather than solving}

%what we actually do in the paper
\medskip
Technically, we proceed as follows. We introduce \emph{monoid network games} (Definition~\ref{def:monoidnetworkgame}) that make common structure of all of our motivating examples explicit, and that we believe cover the majority of network games studied in the literature. Roughly speaking, monoid network games are parametrised wrt \textit{(i)} a monoid that aggregates information from neighbours and \textit{(ii)} functions that govern how that information is propagated in the network. While we are able to model all network games, the structure of monoid network games allows us to characterise them as functors in a generic fashion. %We argue that this is not a too restrictive assumption as it corresponds to the idea that the information that players receive is already aggregated. This is the case in most of economic applications.  \todo{explained that monoid network games are common (maybe check that it is true and cite something in support?) and that we can talk about generic ones as well}

Our category of open graphs $\Graph$ (Definition~\ref{def:Graph}) is an extension of the approach of~\cite{chantawibul_sobocinski_towards_compositional_graph_theory}, from undirected graphs to undirected \emph{multi}graphs.
%\todo{this generalisation is useful}.
Multigraphs allow us to model games on networks where some links are stronger than others, cf. Example~\ref{ex_f:majority}. Our $\Graph$ is different from other notions of ``open graph'' in the literature, e.g.\ via cospans~\cite{Fong2015a}, in that it is centred on the use of \emph{adjacency matrices}, which are commonly used in graph theory to encode connectivity.
%\todo{explained why adjacency matrices are good}.
Adjacency matrices give an explicit presentation of the graphs that allows an explicit description of the games played on them.
Moreover, the emphasis on the matrix algebra means that $\Graph$ has the structure of commutative bialgebra---equivalent to the algebra of ordinary $\mathbb{N}$ matrices~\cite{Lack2004a,Zanasi2015}---but also additional structure that captures the algebraic content of adjacency matrices.
Given that $\Graph$ has a presentation in terms of generators and equations, to obtain~\eqref{eq:Fn} it suffices to define it on the generators and check that  $\Graph$-equations are respected in $\Game$. This is our main result, Theorem~\ref{th:Fn_monoidal_functor}.

In addition to the presentation of $\Graph$ in terms of generators and equations, we characterise it as another category (Theorem~\ref{th:graph_iso_A}) that makes clear its status as a category of ``open graphs''. The result can be understood as a kind of normal form for the morphisms of $\Graph$, useful to describe concrete instantiations of $F_\N$ for arbitrary open graphs (Theorem~\ref{th:game_on_graph}).

\medskip
% motivational paragraph
Our work is a first step towards a more principled way of defining games parametrised by graphs.
%\todo{the concetual approach is more general and it would work with directed graphs or stochastic games as well}.
We would like to remark that the methodology that we present to define games on networks is more general than the particular instance worked out in this paper. Indeed, future work will extend both the notions of graphs (e.g.\ by considering directed graphs), as well as the kinds of games played on them (e.g. \ stochastic games, repeated games). While we do identify some applications, we believe that compositional reasoning is severely under-rated in traditional game theory, and that its adoption will  lead to both more flexible modelling frameworks, as well as more scalable mathematical analyses.

\paragraph*{Structure of the paper.}
%The contents of the paper are as follows. \textit{(i)}
%
We introduce our %network game
running examples in \S\ref{sec:games_on_graphs} and unify them under the umbrella of monoid network games.
Next, we recall the basics of open games in \S\ref{sec:opengames} and identify the building blocks needed for~\eqref{eq:Fn}. In \S\ref{sec:graphs} we introduce the category $\Graph$ of open, undirected multigraphs, and give a combinatorial characterisation, which is useful in applications. The construction of $F_\N$ is in \S\ref{sec:def_functor}, and several applications of our compositional framework are given in \S\ref{sec:example}.

%After giving, in
%Sections~\ref{sec:games_on_graphs} and~\ref{sec:background}, the necessary background about network games, open games and open graphs,
%we define, in Section~\ref{sec:def_functor}, for every \emph{monoid network
%  game} $\N$, a functor $\F \colon \Graph \to \Game$ by indicating its image on
%the generators of $\Graph$. \textit{(ii)} In Section~\ref{sec:exp_functor}, we
%give a concrete form for $\F$ by explicitly specifying its image on a generic
%morphism in $\Graph$\doubt{explain why this is useful}. In order to do this,
%in Section~\ref{sec:combin_graph}, we give a combinatorial presentation of the
%prop $\Graph$. \textit{(iii)} Section~\ref{sec:example} is devoted to fully work
%out some concrete examples using the theory developed in the paper.

\section{Network games}~\label{sec:games_on_graphs}

In this section we introduce motivating examples for our compositional framework and introduce a notion of game called the \emph{monoid network game} that unifies them.

\smallskip
%We refer the reader to \cite{jackson_zenou_games_networks} for an introduction to the theory of network games. Basic definitions and examples are reproduced in this section.

Network games~\cite{bramoulle_handbook_economics_network, jackson_zenou_games_networks} are parametric wrt a  network topology, usually represented by a graph.
Players are the vertices, and the possible connections between the players are represented by the edges.
Moreover, each player's payoff is affected only by the choices of its immediate \emph{neighbours} on the graph.
%In this sense, network games are \emph{games played on graphs}.
We use undirected multigraphs to model network topologies.
%\todo{emphasized that the idea is that the graph represents which players can influence each other}

\begin{definition}\label{def:multigraph}
  An undirected multigraph is
  $G = (V_{G}, E_{G})$, where $V_{G}$ is the
  set of vertices and $E_{G}$ is a sym.\ multi-relation
  on $V_{G}$: a function
  $E_{G} \colon V_{G} \times V_{G} \to \mathbb{N}$
  st $E_{G}(v_{i},v_{j}) = E_{G}(v_{j},v_{i})$.
\end{definition}

A common way of capturing the connectivity of a graph
is via \emph{adjacency matrices}, which play an important role in
graph theory. They are also  crucial for our compositional account.

Assuming an ordering on the set of vertices of a graph,
square matrices $A$ with entries from $\mathbb{N}$ can record connections between vertex $i$ and $j$ in $A_{ij}$: a $0$-entry signifies no edge,
and non-zero entries count the connections.
%  a $1$-entry
% signifies a single edge and entries $>1$ can be used to represent
% multiple connections (in the case of undirected multigraphs).
%
Ordinary matrices are too concrete to
uniquely represent connectivity since edges between $i$ and $j$ can be
recorded in the $(i,j)$th entry or the
$(j,i)$th entry. One could use symmetric matrices or triangular matrices. For us, it is better to equate matrices that encode the same connectivity: $A\sim A'$ iff $A+A^T = A'+{A'}^T$.
\begin{definition}~\label{def:adjacency}
  An \emph{adjacency matrix} is an equivalence class \([A]\) of matrices with entries in the natural numbers. The equivalence relation is given by
  \[A\sim A' \iff A+A^T = A'+{A'}^T.\]
  %An undirected multigraph can, equivalently, be given by a pair
  %$(k_{G},[E_{G}])$, where $k_{G} \in \mathbb{N}$, $E_{G}$ is
  %a $k_{G} \times k_{G}$ matrix of natural numbers and $[ \ ]$
  %indicates the equivalence relation given by $[A] = [A']$ iff
  %$A+A^{T} = A'+A'^{T}$. The matrix $E_{G}$ is usually called
  %adjacency matrix and it records the connectivity of the
  %vertices, namely, $E_{G}(i,j)+E_{G}(j,i)$ is the number of
  %edges between the vertices $i$ and $j$.
\end{definition}
A finite multigraph can also be defined as $(k_{G},[A])$ where $k_{G}\in \mathbb{N}$ and $[A]$ a $k_{G}\times k_{G}$ adjacency matrix.
Let $\bG (n)$ be the set of multigraphs with $n$
vertices, enumerated as $v_1, \ldots, v_n$.

\begin{definition}[Network game]
  An $n$-player \emph{network game} $\N$ consists of, for each player $1 \leq i \leq n$, a set of choices $X_i$ and a payoff $u_i : \bG (n) \times \prod_{j = 1}^n X_j \to \R$, such that each player's payoff is affected only by its own and its neighbours' choices:
%  ---alternative 1---\\
%  for each $G \in \bG (n)$, each player \(i\) and each \(\vect{x},\vect{x}' \in \prod_{j = 1}^n X_j\)
%  \[(x_{i} = x'_{i} \land \forall (v_{i},v_{j}) \in E_{G} \ x_{j}=x'_{j}) \Rightarrow
%  u_i (G, \vect{x}) = u_i (G, \vect{x}')\]
%  ---alternative 2--- (I think this was the reflexive completion one)\\
  for each \(G \in \bG(n)\), each player \(i\), each \(j \neq i\) such that \((v_{i},v_{j}) \notin E_{G}\), each \(\vect{x}_{-j} \in \prod_{k \neq j}^n X_k\), and each \(x_{j},x_{j}' \in X_{j}\)
  \[u_{i}(G, x_{j},\vect{x}_{-j}) = u_{i}(G, x'_{j},\vect{x}_{-j})\]
  (The notation $x_{-j}$, standard in game theory, means a tuple with the $j$th element missing.)\\
  The \emph{set of strategies} is $\prod_{i = 1}^n X_i$ and its elements $\vect{x} \in \prod_{i = 1}^n X_i$ are \emph{strategy profiles}.\\
  The \emph{best response}, for a graph $G \in \bG(n)$, is a relation $\Bf_{\N}$ on the set of strategies, defined by
  $$(\vect{x}, \vect{x}') \in \Bf_{\N} \Leftrightarrow \forall 1\leq i\leq n. \; \forall y_i \in X_i. \; u_i (G, \vect{x} [i \mapsto x_i']) \geq u_i(G, \vect{x} [i \mapsto y_i])$$
  A \emph{Nash equilibrium}, for $G \in \bG (n)$, is a strategy profile $\vect{x}$ s.t.\ for each player $1 \leq i \leq n$, $u_i (G, \vect{x}) \geq u_i (G, \vect{x} [i \mapsto x_i'])$
  for each $x_i' \in X_i$. It is a fix-point of the best response relation.
\end{definition}

We now recall three important examples of network games.

\begin{example}[Majority game]~\label{ex:majority}
  %The majority game is a
  %network game in which
  Each player has two
  choices, $X_i = \{ Y, N \}$.
  A player receives a utility of $1$ if its choice is the majority choice of its neighbours, and $0$ otherwise, i.e.
  \[ u_i (G, \vect{x}) = \begin{cases}
	  1 &\text{ if } | \{ v_j \mid (v_i, v_j) \in E_G \text{ and } x_i = x_j \} |

	  \geq | \{ v_j \mid (v_i, v_j) \in E_G \text{ and } x_i \neq x_j \} | \\
	  0 &\text{ otherwise.}
	\end{cases} \]
  Nash equilibria are strategy profiles where players take the majority choice of their neighbours.
\end{example}

\begin{example}[Best-shot public goods game] \label{ex:best-shot_public_goods}
  Each player has two choices, $X_i = \{ Y, N \}$, interpreted as investing or not investing in a public good.
  The investor bears a cost $0 < c < 1$, and gives a utility of $1$ to themselves and every neighbour.
  % We can imagine that
  The players are already partially satisfied with the current situation and assign a utility of $1-c+\epsilon$, with $0 < \epsilon < c$, to the situation where neither the player nor its neighbours invest.
  The utility functions thus are:
  \[ u_i (G, \vect{x}) = \begin{cases}
	  1 - c &\text{ if } x_i = Y \\
	  1 &\text{ if } x_i = N \text{ and } x_j = Y \text{ for some } (v_i, v_j) \in E_G \\
	  1-c+\epsilon &\text{ otherwise.}
	\end{cases} \]
  The Nash equilibrium is when no player invests, an example of a `tragedy of the commons'.
\end{example}

\begin{example}[Weakest-link public goods game]~\label{ex:weakest-link_public_goods}
  Each player's choice is an investment, valued in $\R_+$.
  The cost to the player given by an increasing cost function $c : \R_+ \to \R_+$ where $c(0)=0$, and utility is the \emph{minimum} level of investment of the player and all neighbours:
  \[ u_i (G,  \vect{x}) = \min_{j = i \text{ or } (v_i, v_j) \in E_G} x_j - c (x_i). \]
  A necessary condition for Nash equilibrium is that no player invests more than its neighbours.
\end{example}

In Examples~\ref{ex:majority},~\ref{ex:best-shot_public_goods} and~\ref{ex:weakest-link_public_goods} every player has the same set of choices, and
the utility depends in a uniform way on neighbours' choices.
We collect these, and other examples in the literature, under the umbrella of \emph{monoid
  network games}.
Most examples in the literature can be collected in two
classes~\cite[ch. 5]{bramoulle_handbook_economics_network},
namely games on networks with constrained continuous actions or
with binary actions. Provided that weights are natural numbers,
the latter can be expressed as monoid network games. To
express the former as monoid network games, we need to
additionally ask that the parameters appearing in the utility
functions of the players be constant. However, we can still
express games of this class with different parameters for
different players by composing different monoid network games.
This is shown in Example~\ref{ex_f:weakest-link_public_goods}.

Network games that do not fall into this category can be nevertheless expressed in a compositional way as illustrated in Fig.~\ref{fig:generic_network_game}.
If a game can be described in the form of a
a \emph{monoid} network game, we can say more:
such games are a monoidal functor from
the category of syntax to the category of semantics.
The details are in Section~\ref{sec:def_functor}.

To the best of our knowledge, the following has not previously appeared in the literature.

\begin{definition}[Monoid network game]\label{def:monoidnetworkgame}
  A \emph{monoid network game} is $\N = (X, M, f, g)$ where:
  \begin{itemize}
	\item $X$ is the set of choices for each player
	\item $M = (M, \oplus, e)$ is a commutative monoid
	\item $f : X \to M$ and $g : X \times M \to \R$ are functions such that each utility function has the form
	  \[ u_i (G,  \vect{x}) = g \left(
	  x_i, \bigoplus_{(v_i, v_j) \in E_G} f (x_j) \right).
	  \]
  \end{itemize}
\end{definition}

Examples~\ref{ex:majority},~\ref{ex:best-shot_public_goods},~\ref{ex:weakest-link_public_goods} are indeed examples of monoid network
games:
%\begin{example}[Majority game]~\label{ex_m:majority}
\begin{itemize}
\item~\label{ex_m:majority}
  The majority game (Example~\ref{ex:majority}) has the monoid $(\mathbb N, +, 0) \times (\mathbb N, +, 0)$, counting the $Y$ and $N$ `votes'.
  Define $f : \{ Y, N \} \to \mathbb N^2$ by $f (Y) = (1, 0)$ and $f (N) = (0, 1)$, and $g : \{Y, N \} \times \mathbb N^2 \to \R$ is:
  \[ g (x, (n_1, n_2)) = \begin{cases}
	  1 &\text{ if } x = Y \text{ and } n_1 \geq n_2 \\
	  1 &\text{ if } x = N \text{ and } n_1 \leq n_2 \\
	  0 &\text{ otherwise.}
	\end{cases} \]
%\end{example}
%\begin{example}[Best-shot public goods game]~\label{ex_m:best-shot_public_goods}
\item~\label{ex_m:best-shot_public_goods}
  The best-shot public goods game (Example~\ref{ex:best-shot_public_goods}) is a monoid network game with
  the monoid $\mathbf{Bool} = (\{Y, N \}, \vee, N)$, where
  $\vee$ is logical or,
  %  $N \vee N = N$ and $x \vee y = Y$ otherwise.
  % The set of choices
  %for every player is $\{Y, N\}$ as well.
  $f : \mathbf{Bool} \to \mathbf{Bool}$ is the identity, and $g : \mathbf{Bool} \times \mathbf{Bool} \to \R$:
  \[ g (x, y) = \begin{cases}
	  1 - c &\text{ if } x = Y \\
	  1 &\text{ if } x = N \text{ and } y = Y \\
	  1-c+\epsilon &\text{ if } x = N \text{ and } y = N
	\end{cases} \]
%\end{example}
\item~\label{ex_m:weakest-link_public_goods}
%\begin{example}[Weakest-link public goods game]~\label{ex_m:weakest-link_public_goods}
  The weakest-link public goods game (Example~\ref{ex:weakest-link_public_goods}) %is a monoid network game with the
  has the monoid
   $\R_+^\infty = (\{ \R_+ \cup \{ \infty \}, \min, \infty)$,
   $f$ the embedding $\R_+ \hookrightarrow \R_+^\infty$, and $g : \R_+ \times \R_+^\infty \to \R$ is $g (x, y) = \min (x, y) - c (x)$.
%\end{example}
\end{itemize}

\section{Open games}~\label{sec:opengames}

Open games were introduced in~\cite{hedges_etal_compositional_game_theory} as a compositional approach to game theory.

\begin{definition}[Open game] Let $X, Y, R, S, \Sigma$ be sets.
An open game $\game{\mathcal{G}}{X}{S}{Y}{R}{\Sigma}$ has:
  \begin{enumerate}[(i)]
    \item $\Pf_{\mathcal{G}} \colon \Sigma \times X \to Y$,
called play function
    \item
$\Cf_{\mathcal{G}} \colon \Sigma \times X \times R \to S$, called
coplay function
    \item
$\Bf_{\mathcal{G}} \colon X \times (Y \to R) \to \pset(\Sigma^2)$,
called best response function.
  \end{enumerate}
\end{definition}

Roughly speaking, an open game is a process that \textit{(i)}
given a \emph{strategy} and \emph{observation}, decides a \emph{move}, and
\textit{(ii)} given a strategy, observation, and a \emph{utility},
returns a \emph{coutility} to the environment. Coutility is not a
concept of classical game theory, but it enables
compositionality by incorporating the fact that
players reason about the future consequences of their
actions. Finally, \textit{(iii)}, the best response function,
which, given a context for the game returns a relation on the set
of strategies. A strategy $\sigma$ is related to another strategy
$\sigma'$ if the latter is a best response to the former.

An open game is %thought of as
a process that receives observations $(X)$
  \emph{`from the past'}, and the utility $(R)$ \emph{`from the
future'}. It outputs
 moves $(Y)$ covariantly and coutility $(S)$ contravariantly.
\begin{center}\genericopengame\end{center}
Open games are morphisms in a symmetric monoidal category $\Game$.
In order to formally define composition and monoidal product of games, it
is useful to rephrase the definition in terms of
lenses~\cite{oles82}. The detailed definitions are given in~\cite{hedges_etal_compositional_game_theory}.
% The details are recalled in Appendix~\ref{app:opengames}.
\begin{definition}[$\Game$] $\Game$ is the symmetric
monoidal category with pairs of sets $\diset{X}{S}$ as objects and
(equivalence classes of) open games
$\game{\mathcal{G}}{X}{S}{Y}{R}{\Sigma}$ as morphisms.
\end{definition}

We give some intuitions.
Composition, shown below left, is sequential play:
$\mathcal{H} \cdot \G$ is thought of as $\mathcal{H}$
happening after $\G$, observing the moves of $\G$ and feeding
back its coutility as $\G$'s utility.
The monoidal product of open games represents two games played
independently. The games are placed
 side by side with no connections, as shown below right.
\begin{center}\compositionopengames \qquad\qquad \tensoropengames\end{center}

Classical games are
\emph{scalars} in $\Game$, i.e.\ open games
$\diset{1}{1}\nrightarrow\diset{1}{1}$. The fix-points of the
best response functions of scalars in $\Game$ are the Nash
equilibria of the games they represent.

\medskip

Next we define specific open games used in
our compositional account of network games. The first
is the \textsf{Utility Maximising Player}, modelling typical players of classical game theory.

\begin{definition}[Utility Maximising Player] Let $X$ and $Y$ be
sets and
$\mathsf{argmax} \colon {\R}^Y \to \mathcal{P}(Y)$
take a function $\cont :Y \to \R$
to the subset of $Y$ where $\cont$ is maximised.
Define $\mathcal{D}$ to be:\\
  \begin{minipage}{0.6 \textwidth}
    \begin{align*} & \game{\mathcal{D}}{X}{1}{Y}{\R}{Y^X}\\ &\begin{cases} \Pf_{\mathcal{D}}(f,x) = f(x)\\ \Cf_{\mathcal{D}}(f,x,r) = *\\ \Bf_{\mathcal{D}}(x,\cont) = \lbrace (y,y') \in Y^X\times Y^X : y'(x) \in \mathsf{argmax}(\cont) \rbrace
      \end{cases}
    \end{align*}
  \end{minipage}
  \begin{minipage}{0.4 \textwidth}
    \[ \utilitymaximisingplayer \]
  \end{minipage}
\end{definition}

The category of sets and functions $\mathbf{Set}$ embeds into
$\Game$ in two ways. In our account of network
games, these embeddings encode how neighbours influence each
other's utilities.

\begin{definition}\label{def:lifted_functions} Let $X, Y$ be sets and $f \colon X \to Y$ a
function. Its covariant lifting is defined:\\
  \begin{minipage}{0.6 \textwidth}
    \begin{align*} & \game{f^*}{X}{1}{Y}{1}{1}\\ &\begin{cases} \Pf_{f^*}(*,x) = f(x)\\ \Cf_{f^*}(*,x,*) = *\\ \Bf_{f^*}(x,*) = \lbrace (*,*) \rbrace
      \end{cases}
    \end{align*}
  \end{minipage}
  \begin{minipage}{0.4 \textwidth}
    \[ \liftingopengame \]
  \end{minipage}\\

Similarly, its contravariant lifting is the
following:\\
  \begin{minipage}{0.6 \textwidth}
    \begin{align*} & \game{f_*}{1}{Y}{1}{X}{1}\\ &\begin{cases} \Pf_{f_*}(*,*) = *\\ \Cf_{f_*}(*,*,x) = f(x)\\ \Bf_{f_*}(*,x) = \lbrace (*,*) \rbrace
      \end{cases}
    \end{align*}
  \end{minipage}
  \begin{minipage}{0.4 \textwidth}
    \[ \coliftingopengame \]
  \end{minipage}
\end{definition}

\begin{figure}
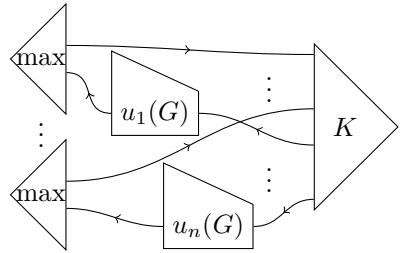

  \centering
  \genericnetworkgameutility
  \caption{Open game representing a network game $\N$ played on a
    multigraph $G$}
  \label{fig:generic_network_game}
\end{figure}

To obtain Examples~\ref{ex:majority},~\ref{ex:best-shot_public_goods}
and~\ref{ex:weakest-link_public_goods} as scalars in $\Game$,
players are taken to be utility-maximising players.
The connectivity of the multigraph $G$ determines
their utility functions as contravariant
liftings $u_i(G)$, while the
context $K$ sends back the choices of all players:
\begin{align*}
  & \game{K}{X^n}{X^n \times \cdots \times X^n}{1}{1}{1} \\
  & \Cf_{K}(\vect{x}) = (\vect{x},\ldots ,\vect{x}).
\end{align*}
The respective games are then obtained as
the composition illustrated in Fig.~\ref{fig:generic_network_game}.
In this way, we obtain a compositional description of any network game.
If a game can be described in the form of a
a \emph{monoid} network game, we can say more:
such games are a monoidal functor from
$\Graph$, defined in the next section, to $\Game$.
The details are in Section~\ref{sec:def_functor}.

\section{Open graphs}~\label{sec:graphs}

We extend the %algebraic approach to
compositional approach to
graph theory
of~\cite{chantawibul_sobocinski_towards_compositional_graph_theory}
from simple graphs to undirected multigraphs, identifying a
``syntax'' of network games as the arrows of a prop\footnote{A prop~\cite{MacLane1965,Lack2004a} is a symmetric strict monoidal category where the objects are $\mathbb{N}$, and $m\otimes n := m+n$.}
$\Graph$, generated from a monoidal signature and equations. We also
 provide a characterisation of $\Graph$ that explains its arrows as ``open graphs''.
Differently from other approaches~\cite{Bonchi2016,Fong2015a},
$\Graph$ uses adjacency matrices (Definition~\ref{def:adjacency}).
Indeed, the presentation includes generators
\begin{equation}\label{eq:bialggen}\tag{BIALG}
  $\graphgenerators$
\end{equation}
and the equations of
Fig.~\ref{fig:commutative-bialgebra}.
The prop $\bialg$ generated
by this data is isomorphic~\cite{Lack2004a,Zanasi2015} to the prop of matrices with entries
from $\mathbb{N}$, with composition being matrix
multiplication. To convert between the two, think of
  the matrix as recording the numbers of paths: indeed, the $(i,j)$th
  entry in the matrix is the number of paths from the $i$th left port to the $j$th right port.
\begin{example}\label{ex:diagram_matrix}
  The following string diagram in $\bialg$
	corresponds to the $3\times 2$ matrix
  $\left(\begin{smallmatrix}2 & 1\\0 & 1\\1 & 0 \end{smallmatrix}\right)$.
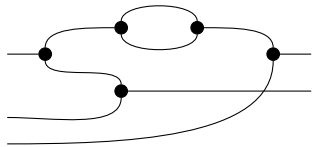
  \begin{center}
    \exDiagramMatrix
  \end{center}
\end{example}
\begin{figure*}
  \begin{center}\commutativebialgebraaxioms\end{center}
  \caption{Commutative bialgebra equations, yielding prop $\bialg$.}
  \label{fig:commutative-bialgebra}
\end{figure*}
\begin{figure*}
  \begin{center}\cupequations\end{center}
  \caption{Equations of $\cup$, which together with the equations of Fig.~\ref{fig:commutative-bialgebra} yield prop $\hau$.}
  \label{fig:cup-equations}
\end{figure*}
Next, we add a ``cup'' generator denoted
\begin{equation}\label{eq:cupgen}\tag{U}
  \cupgenerator
\end{equation} with its equations given in
Fig.~\ref{fig:cup-equations}.
\begin{definition}
Let $\hau$ be the prop obtained
from~\eqref{eq:bialggen} and~\eqref{eq:cupgen},
quotiented by equations in Figs.~\ref{fig:commutative-bialgebra}
and~\ref{fig:cup-equations}, where the empty diagram is the identity on the monoidal unit.
\end{definition}
Just as $\bialg$ captures ordinary
matrices, $\hau$ captures adjacency
matrices:
\begin{proposition}\label{pro:abstractadjacency}
  For $n\in\mathbb{N}$, the hom-set $[n,0]$ of $\hau$ is in bijection
  with  $n\times n$ adjacency matrices, in the sense of Definition~\ref{def:adjacency}.
\end{proposition}
We have seen that the relationship between matrices and diagrams in $\bialg$ is that the former encode the path information from the latter. Thus an $m\times n$ matrix is a diagram from $m$ to $n$. Adding the cup and the additional equations means that, in general, a diagram from $n$ to $0$ in $\hau$ ``encapsulates'' an $n\times n$ matrix that expresses connectivity information in a similar way to adjacency matrices. We now give a concrete derivation to demonstrate this.
\begin{example}\label{ex:abstractadjacency}
  The equivalence relation of adjacency matrices is captured by
  the equations of Fig.~\ref{fig:cup-equations}.
  Consider matrices
  $A = \left(\begin{smallmatrix}0 & 1\\1 & 0\end{smallmatrix}\right) \sim \left(\begin{smallmatrix}0 & 2\\0 & 0\end{smallmatrix}\right) = A'$.
  The morphism in $\hau$ is obtained by
  constructing their diagram in $\bialg$ as in
  Example~\ref{ex:diagram_matrix} and ``plugging'' them in the
  following.
  \begin{center}
    \diagrameqrelationUmat
  \end{center}

  As shown below, the two diagrams obtained are equated by the axioms of $\hau$.

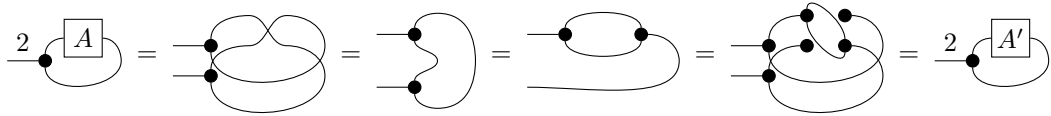
  \[\exDiagramEquivalenceAdjacencyFirst \ = \ \exDiagramEquivalenceAdjacencySecond \ = \ \exDiagramEquivalenceAdjacencyThird \ = \ \exDiagramEquivalenceAdjacencyFourth \ = \ \exDiagramEquivalenceAdjacencyFifth \ = \ \exDiagramEquivalenceAdjacencySixth \]

\end{example}
The prop $\hau$ can be given a straightforward combinatorial characterisation as the prop $\Umat$.
\begin{definition}[$\Umat$]\label{def:Umat}
  A morphism $\alpha \colon m \to n$ in the prop
  $\Umat$~\cite{chantawibul_sobocinski_towards_compositional_graph_theory}
  is a pair $(B,[A])$, where $B \in \Mat_{\mathbb{N}}(m,n)$ is a
  matrix, while $[A]$, with $A \in \Mat_{\mathbb{N}}(m,m)$,
  is an adjacency matrix.
  The components of $\Umat$ morphisms can be read off a
  ``normal form'' for $\hau$ arrows, as follows.
  \[ \defpropUmat \]
\end{definition}

Composition in $\Umat$ becomes intuitive when visualised with
string diagrams.
\[ (B,[A]) \circ (B',[A']) = (BB',[A+BA'B^{T}])\]
\defcompositionUmat

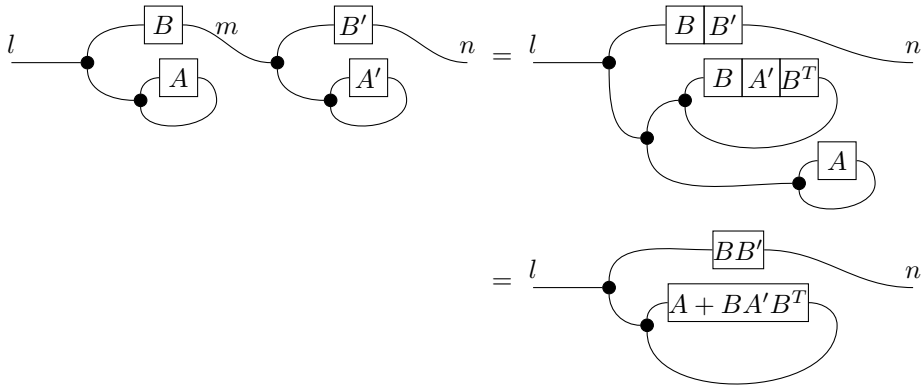
\begin{proposition}\label{pro:hauiso}
  $\hau$ is isomorphic to the prop  $\Umat$. \qed
\end{proposition}
The proof
is similar to the case for $\mathbb{Z}_2$~\cite{chantawibul_sobocinski_towards_compositional_graph_theory}.
An extension of $\hau$ with just one additional generator and no
additional equations yields the prop $\Graph$ of central interest for us.
\begin{definition}\label{def:Graph} The prop $\Graph$ is obtained from the
  generators in~\eqref{eq:bialggen} and~\eqref{eq:cupgen} together
  with a generator $\vertexgenerator$.
	The equations are those of
  Figs.~\ref{fig:commutative-bialgebra}
  and~\ref{fig:cup-equations}.
\end{definition}
As we shall see, arrows $0\to 0$ in $\Graph$ are precisely finite
undirected multigraphs taken up to isomorphism: the additional generator
plays the role of a graph vertex.
\begin{example}\label{ex:closed_graphs}
  For example, the first of the following represents a multigraph with
  two vertices, connected by a single edge. The second one, two
  vertices connected by two edges. The third one, is a multigraph
  with three vertices and two edges between them.
  \begin{center}
    \exClosedgraphs
  \end{center}
\end{example}
While the arrows $[0,0]$ are (iso classes of)
multigraphs, general arrows can be understood as open graphs. Roughly speaking, they are
graphs together with interfaces, and data that specifies the
connectivity of the graph to its interfaces. We make this explicit below. Indeed, we shall see (Theorem~\ref{th:graph_iso_A}) that the prop $\A$, defined below, is isomorphic to $\Graph$ -- for this reason we use $\Graph$ string diagrams to illustrate its structure.
\begin{definition}[The prop $\A$]~\label{def:A} A morphism
  $\Gamma \colon m \to n$ in the prop $\A$ is defined by
  \begin{equation}\label{eq:A} \Gamma = (k, \left[ A \right] , B, C, D, \left[ E \right])
  \end{equation} where $k \in \mathbb{N} $,
  $A \in \Mat _{\mathbb{N}}(m,m)$, $B \in \Mat _{\mathbb{N}}(m,n)$,
  $C \in \Mat _{\mathbb{N}}(m,k)$, $D \in \Mat _{\mathbb{N}}(k,n)$
  and $E \in \Mat _{\mathbb{N}}(k,k)$.
	Similarly to $\Umat$ (Definition~\ref{def:Umat}),
	the components of~\eqref{eq:A} can be read
  off a ``normal form'' for arrows of $\Graph$, as visualised below right.\\

	\begin{minipage}{0.5 \textwidth}
  Tuples~\eqref{eq:A} are taken up to an equivalence
  relation that captures the fact that the order of the vertices is immaterial. Let $\Gamma \sim \Gamma'$ iff they are morphisms of the same type,
  $\Gamma , \Gamma' \colon m \to n$ with $k$ vertices, and there is
  a permutation matrix $P \in \Mat(k,k)$ such that
  $\Gamma' = (k, \left[ A \right], B, CP^T, PD, \left[ PEP^T \right])$.
  The justification for this equivalence is
  the equality of the following two string diagrams in $\Graph$,
  below (for the details, see Appendix~\ref{app:graphs} on page~\pageref{details_equivalence_A}).
	\end{minipage}
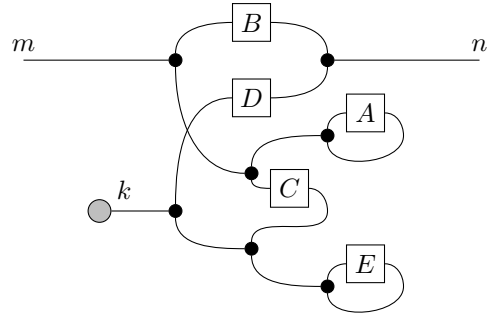
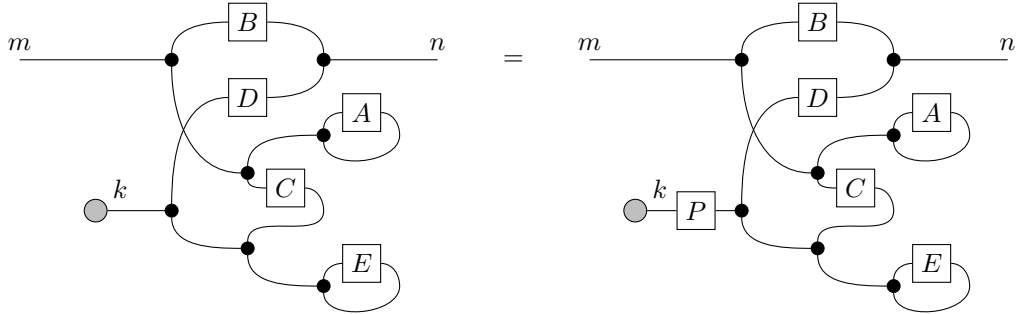
	\begin{minipage}{0.5 \textwidth}
	\[ \defpropA \]
	\end{minipage}
	\begin{center}
    \defeqrelationA
	\end{center}
\end{definition}
It is worthwhile to give some intuition for the
components of~\eqref{eq:A}. The idea is that an arrow $\Gamma$
specifies a multigraph $G=(k,[E])$, and:
\begin{itemize}
  \item $B$ specifies connections between the two boundaries,
    bypassing $G$
  \item $C$ specifies connections between the left boundary
    and $G$
  \item $D$ specifies connections between $G$ and the right
    boundary
  \item $A$ specifies connections between the interfaces on
    the left boundary. This allows $\Gamma$ to introduce connections
    between the vertices of an ``earlier'' open graph $\Delta$. See Example~\ref{ex:compositioninA}.
\end{itemize}

Defining composition in $\A$ is straightforward, given the above intuitions, but
the details are rather tedious: see Lemma~\ref{lem:aprop} in Appendix~\ref{app:graphs}.

\begin{example}\label{ex:compositioninA} In a composite
  $\Delta \mathrel{;} \Gamma$, $\Gamma$ may introduce
  edges between the vertices of $\Delta$. Indeed, the
  first diagram in Example~\ref{ex:closed_graphs} can be
  decomposed:
  %, where the second component connects the two vertices in the first component.
  \begin{center}
    \exCompositionOpenGraphs
  \end{center}
\end{example}
\begin{example}
  The following show the role of $\A$-morphism components, when isolated.
  % as the only non-trivial component.
  The leftmost open graph has only left-side ports. It
  introduces a self-loop and two connections.
  The second has only connections between the left and right
  interfaces; the first left port is connected twice to the
  first right port, the second port is disconnected, and the third left port is connected to the second and third right ports. The third open graph
  has one vertex connected to the two left ports.
  The fourth has three vertices connected to the right
  ports, following the specification in the second. The rightmost (closed) multigraph has its vertices
  connected according to the specification of the
  leftmost vertex-less open graph. We write $!$ for matrices without columns, \udexcl~for matrices without rows and $( \ )$ for the empty matrix.
  \exIsolateComponentsGraphs
\end{example}
\medskip The main result in this section is the following.
\begin{theorem}\label{th:graph_iso_A}
  There is an isomorphism of props $\isograph \colon \Graph \to \A$.
\end{theorem}
\begin{figure}
\imageisograph
\caption{Image of $\isograph$ on the generators}
\label{fig:iso_graph}
\end{figure}
The remainder of this section builds a proof of the above, summarised in the diagram below.
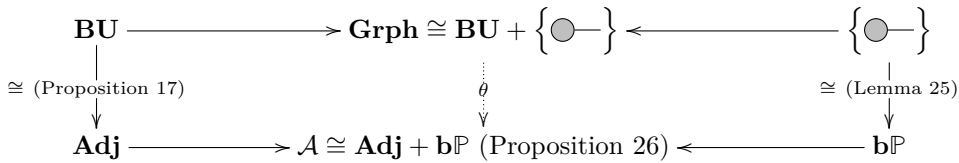
\[
\xymatrix@C=60pt{
\hau \ar[d]|{\cong\ \text{(Proposition~\ref{pro:hauiso})}}\ar[r] & \Graph \cong \hau + \freeononevertex
\ar@{.>}[d]|{\theta}
& \ar[l] \freeononevertex
\ar[d]|{\cong\ \text{(Lemma~\ref{lemma:bij_is_free})}}
\\
\Umat \ar[r] & \A \cong \Umat + \bij\  \text{(Proposition~\ref{pro:Aiscoproduct})} & \ar[l] \bij
}
\]
First, note that
$\Graph$ is the coproduct $\hau + \freeononevertex$ in the
category of props, where
$\freeononevertex$ is the free prop on a single generator $0\rightarrow 1$.
Next, we
characterise $\freeononevertex$ as $\bij$, defined below, in Lemma~\ref{lemma:bij_is_free}.
Given that $\hau \cong \Umat$, as shown in  Proposition~\ref{pro:hauiso}, to show the existence of
$\theta$ it suffices to show that $\A$ satisfies the
universal property of the coproduct $\Umat + \bij$, which is Proposition~\ref{pro:Aiscoproduct}.
The action
of $\theta$ on the generators of $\Graph$ is in Fig.~\ref{fig:iso_graph}.

\begin{definition}[$\bij$]~\label{def:bij}
  The prop of
  \emph{bound permutations} $\bij$ has as morphisms $m \to m+k$
	pairs $[(k,P)]$ where $k\in \mathbb{N}$ and $P \in \Mat _{\mathbb{N}}(m+k,m+k)$ is a permutation matrix.
	Such pairs are identified to ensure that the order of the lower $k$
  rows of $P$ is immaterial. Roughly speaking, considering $P$ as
  a permutation of $m+k$ inputs to $m+k$
  outputs, in $[(k,P)]$ the final
  $k$ inputs are ``bound''.
  Explicitly, $(k,P) \sim (k,P')$
  iff there is a permutation
  $\sigma \in \Mat _{\mathbb{N}}(k,k)$ st
  $P = \left( \begin{smallmatrix} \mathbb{1}_m & \mathbb{0}\\ \mathbb{0} & \sigma \end{smallmatrix} \right) P'$.
  Composition is defined:
  \begin{align*} & (l,Q) \circ (k,P) = (k+l, \left( \begin{smallmatrix} P & \mathbb{0} \\ \mathbb{0} & \mathbb{1}_l \end{smallmatrix} \right) Q) & \defcompositionbij
  \end{align*} Identities are identity matrices $id_n = (0,\mathbb{1}_n)$.
  The fact
  that $\bij$ is a prop is Lemma~\ref{lem:bijprop}
  in Appendix~\ref{app:graphs}.
\end{definition}

\begin{lemma}~\label{lemma:bij_is_free}
  $\bij$ is isomorphic to $\freeononevertex$.
\end{lemma}
\begin{proof} Let us call $\phi = (0, (1)) \colon 0 \to 1$, which
  is a morphism in $\bij$. We show directly that, for any other prop
  $\mathbb{P}$ that contains a morphism $v \colon 0 \to 1$, there
  is a unique prop homomorphism
  $\alpha^{\#} \colon \bij \to \mathbb{P}$ such that
  $\alpha^{\#}( \phi) = v$. The details are given as
  Lemma~\ref{lem:bij_is_free_full} in Appendix~\ref{app:graphs}.
\end{proof}

Given the results of Proposition~\ref{pro:hauiso} and Lemma~\ref{lemma:bij_is_free}, we obtain the isomorphism
 $\theta \colon \Graph \to \A$, thereby completing
 the proof of Theorem~\ref{th:graph_iso_A}, by showing that:
\begin{proposition}\label{pro:Aiscoproduct}
$\A$ satisfies the universal property of the coproduct $\Umat + \bij$.
\end{proposition}
\begin{proof} In order to show that $\A$ is a coproduct
  $\Umat + \bij$, we define the two inclusions.\\
  \begin{minipage}{0.4 \textwidth}
    \begin{align*} i_1 \colon \Umat & \longrightarrow \A\\ n & \longmapsto n \\ (B, \left[ A \right] ) & \longmapsto (0, \left[ A \right], B, !, \text{!`}, \left[ ( \ ) \right] ) \\
       \definclusiononeALHS &\longmapsto \definclusiononeARHS
    \end{align*}
  \end{minipage}
  \begin{minipage}{0.5 \textwidth}
    \begin{align*} i_2 \colon \bij & \longrightarrow \A\\ n & \longmapsto n \\ (k,P) & \longmapsto (k, \left[ \mathbb{0}_n \right], P\row{[1,n]}, \mathbb{0}_{nk}, P\row{[n+1,n+k]}, \left[ \mathbb{0}_k \right] )\\
       \definclusiontwoALHS &\longmapsto \definclusiontwoARHS
    \end{align*}
  \end{minipage}

  We indicate with $P\row{[1,n]}$ the first $n$ rows of the
  matrix $P$ and, similarly, with $P\row{[n+1,n+k]}$ the rows
  between the $n+1$-th and the $n+k$-th.
  It is not difficult to show that these are indeed homomorphism,
  the details are given as
  Claim~\ref{cl:injections-are-homs} in Appendix~\ref{app:graphs}.

  Now, we show that, for any other prop $\mathcal{C}$ with
  prop homomorphisms
  $\Umat \overset{f_1}{\longrightarrow} \mathcal{C} \overset{f_2}{\longleftarrow} \bij$,
  there exists a unique prop homomorphism
  $H \colon \A \to \mathcal{C}$ such that $H \circ i_1 = f_1$ and
  $H \circ i_2 = f_2$.\label{pg:Hdef}
  %Define~\label{pg:Hdef}the map:
  \begin{align*}
    H \colon \A &\longrightarrow \mathcal{C}\\ n & \longmapsto n\\ (k, \left[ A \right] , B, C, D, \left[ E \right]) & \longmapsto f_1 \left( \left( \begin{smallmatrix} B \\ D\end{smallmatrix} \right), \left[ \left( \begin{smallmatrix} A & C \\ \mathbb{0} & E \end{smallmatrix} \right) \right] \right) \circ \left( \mathbb{1}_m \otimes f_2(k, \mathbb{1}_k) \right) \\
  \end{align*}
  We verify that $H$ is a homomorphism in
  Lemma~\ref{lem:Hhomo} in Appendix~\ref{app:graphs}. Next, we
  confirm that $H \circ i_1 = f_1$ and $H \circ i_2 = f_2$, where two functor boxes~\cite{functor-boxes} for $f_{1}$ and $f_{2}$ are coloured:
  \begin{multline*}
    H \circ i_1(B, \left[ A \right]) = H(0, \left[ A \right], B, !, \text{!`}, \left[ ( \ ) \right]) \\
    = f_1(B, \left[ A \right]) \circ (\mathbb{1}_m \otimes f_2(0, \mathbb{1}_0)) \left(\propertyoneHFirst\right) = f_1(B, \left[ A \right]) \left(\propertyoneHSecond\right)
  \end{multline*}
	\begin{multline*}
    H \circ i_2 (k, P) = H(k, \left[ \mathbb{0}_n \right], P\row{[1,n]}, \mathbb{0}_{nk}, P\row{[n+1,n+k]}, \left[ \mathbb{0}_k \right] ) \\
    = f_1(P, \left[ \mathbb{0}_{n+k} \right]) \circ ( \mathbb{1}_n \otimes f_2(k, \mathbb{1}_k)) \left(\propertytwoHFirst\right)
    = P \circ f_2(k, \mathbb{1}_{n+k}) \left(\propertytwoHSecond\right)\\
    = f_2(k, P) \left(\propertytwoHThird\right)
  \end{multline*}

  Moreover, $H$ is the unique prop homomorphism with these
  properties. In fact, suppose there is $H' \colon \A \to C$ such
  that $H' \circ i_1 = f_1$ and $H' \circ i_2 = f_2$. Then:
  \begin{align*}
    H'(k, \left[ A \right], B, C, D, \left[ E \right]) =& H'(i_1( \left( \begin{smallmatrix} B \\ D \end{smallmatrix} \right), \left[ \left( \begin{smallmatrix} A & C \\ \mathbb{0} & E \end{smallmatrix} \right) \right]) \circ (\mathbb{1}_m \otimes i_2(k, \mathbb{1}_k)))\\
    =& H' i_1( \left( \begin{smallmatrix} B \\ D \end{smallmatrix} \right), \left[ \left( \begin{smallmatrix} A & C \\ \mathbb{0} & E \end{smallmatrix} \right) \right]) \circ (H'(\mathbb{1}_m) \otimes H' i_2(k, \mathbb{1}_k))\\
    =& f_1( \left( \begin{smallmatrix} B \\ D \end{smallmatrix} \right), \left[ \left( \begin{smallmatrix} A & C \\ \mathbb{0} & E \end{smallmatrix} \right) \right]) \circ (\mathbb{1}_m \otimes f_2(k, \mathbb{1}_k))
    = H(k, \left[ A \right], B, C, D, \left[ E \right]).
  \end{align*}
  %Therefore, $\A \cong \Umat + \bij \cong \Graph$. The
  %isomorphism $\isograph \colon \Graph \to \A$ can be defined
  %explicitly on the generators of $\Graph$ as in Figure~\ref{fig:iso_graph}.
\end{proof}

\section{Games on graphs via functorial semantics}~\label{sec:def_functor}

Here we show that monoid network games $\N$ define monoidal functors $F_\N \colon \Graph \to \Game$, which is our main contribution.
To every open graph $\Gamma$, $F_\N$ associates an open game, where $\N$ is played on $\Gamma$. We give an explicit account of the $F_\N$-image of open graphs $\Gamma$, using Theorem \ref{th:graph_iso_A}. We also explain how $F_\N$ acts on closed graphs, giving classical games.
\smallskip

%, we define a
%monoidal functor $F_\N : \Graph \to \Game$.\\
%
Since $\Graph$ is given by generators and equations, it suffices to define $F_\N$ on the generators and show that the equations are respected.
Fix a monoid network game $\N = (X,M,f,g)$. % (Definition~\ref{def:monoidnetworkgame}).
\begin{itemize}
	\item On objects, $F_\N (1) = \diset{M}{M}$. Thus, for $n \in \Graph$, we have $F_\N (n) = \diset{M^n}{M^n}$
	\item The vertex $\tinyvertex : 0 \to 1$ is mapped
	to the open game $\game{F_\N (\tinyvertex)}{1}{1}{M}{M}{X}$
	defined\\
	\begin{minipage}{.5\textwidth}
	\begin{itemize}
		\item $\Sigma_{F_\N (\microvertex)} = X$
		\item $\Pf_{F_\N (\microvertex)} (x_i, *) = f (x_i)$
		\item $\Cf_{F_\N (\microvertex)} (x_i, *, m) = *$
      \item $\begin{multlined}[t](x_i, x_i') \in \Bf_{F_\N (\microvertex)} (*, \cont : M \to M) \\
        \text{ iff } x_i' \in \arg\max_{x_i'' : X} g (x_i'', \cont (f (x_i'')))\end{multlined}$
	\end{itemize}
    \end{minipage}
	\begin{minipage}{.4\textwidth}
	  \begin{center}
		\imagefunctorvertex
	  \end{center}
	\end{minipage}
  \item The generators~\eqref{eq:bialggen} are mapped to the bialgebra structure on $(M, M)$ induced by
	 the monoid action of $M$. Specifically, they are:
	\begin{center}
      \imagefunctorbialgebra
    \end{center}
    where each of these open games is built from lifted functions
	(Definition~\ref{def:lifted_functions}).

  \item $\tinycup : 2 \to 0$ is mapped to the following open game (see \cite{hedges_etal_compositional_game_theory})\\
	\begin{minipage}{.5\textwidth}
	  \begin{align*}
		&\game{F_\N (\tinycup)}{M^{2}}{M^{2}}{1}{1}{1}\\
		&\begin{cases}
		  \Pf(*,m_{1},m_{2})=*\\
		  \Cf(*,m_{1},m_{2},*)=(m_{2},m_{1})
		\end{cases}
	  \end{align*}
    \end{minipage}
	\begin{minipage}{.5\textwidth}
	\begin{center}
      \imagefunctorcup
    \end{center}
	\end{minipage}
\end{itemize}
To prove that $F_\N$ is a symmetric monoidal functor it suffices to show that the equations of $\Graph$ are respected; this is a straightforward but somewhat lengthy computation.
\begin{theorem}~\label{th:Fn_monoidal_functor}
	$F_\N$ defines a symmetric monoidal functor $\Graph \to \Game$.
\end{theorem}
\begin{proof}
See Appendix~\ref{app:def_functor}, on page~\pageref{proof:th_F_functor}.
\end{proof}

Note that $F_\N$ does \emph{not} respect axioms (C1) or (C2) of \cite{chantawibul_sobocinski_towards_compositional_graph_theory}, so it does not define a functor $\ABUV \to \Game$ in the terminology of \emph{loc.\ cit.} This, together with the increased expressivity of multigraphs over simple graphs, motivates our extension from $\ABUV$ to $\Graph$.

%\todo{Note that $F_\N$ does not respect axioms (C1) or (C2), so it does not define a functor $\ABUV \to \Game$.}

 Theorem~\ref{th:graph_iso_A} gives a convenient ``normal form'' for the arrows of $\Graph$, which we use to give an explicit description of the image of any (open) graph $\Gamma$ under $F_\N$.  First, we specialise to closed graphs that yield ordinary network games. This result---a sanity check for our compositional framework---is a corollary of the more general Theorem \ref{th:game_on_graph}, proved subsequently.

\begin{corollary} \label{th:closed_game_on_graph}
	Let $\N =(X,M,f,g)$ be a monoid network game, and consider
	$\Gamma : 0 \to 0$ in $\Graph$, an undirected multigraph with $k$ vertices.
	Then the game $\game{F_\N (\Gamma)}{1}{1}{1}{1}{X^{k}}$ has:
	\begin{itemize}
		\item $\Sigma_{F_\N (\Gamma)} = X^{k}$ as its strategy profiles,
		\item
		$\Bf_{F_\N (\Gamma)} (*, *) \subseteq X^{k} \times X^{k}$ is
		the best response relation of $\N$ played on
	  $\Gamma$.
	\end{itemize}
\end{corollary}
 Note that while the expressions in the statement of Theorem~\ref{th:game_on_graph} below may seem involved, they are actually derived in an entirely principled, compositional manner from the generators of $\Graph$. Indeed, the proof is by structural induction on the morphisms of $\Graph$.

\begin{theorem} \label{th:game_on_graph}
	Let $\mathcal N=(X,M,f,g)$ be a monoid network game.
	Let $\Gamma : i \to j$ be a morphism in $\Graph$ with $k$ vertices st  $\theta(\Gamma)=(k, \left[ A \right], B, C, D, \left[ E \right])$, where $A : i \times i$, $B : i \times j$, $C : i \times k$, $D : k \times j$ and $E : k \times k$.
	Then the open game $\game{F_\mathcal N (\Gamma)}{M^i}{M^i}{M^j}{M^j}{X^{k}}$ has:
	\begin{itemize}
		\item set of strategy profiles $\Sigma (F_\mathcal N (\Gamma)) = X^k$
		\item play function $\Pf_{F_\mathcal N (\Gamma)} : X^k \times M^i \to M^j$ given by
		$\Pf_{F_\mathcal N (\Gamma)} (\vect{\sigma}, \vect{x}) = B^T \vect{x} \oplus D^T f(\vect{\sigma})$
		\item coplay function $\Cf_{F_\mathcal N (\Gamma)} : X^k \times M^i \times M^j \to M^i$ is $\Cf_{F_\mathcal N (\Gamma)} (\vect{\sigma}, \vect{x}, \vect{r}) = (A+A^T)\vect{x} \oplus B\vect{r} \oplus C f(\vect{\sigma})$
		\item best response relation $\Bf_{F_\mathcal N (\Gamma)} : M^i \times (M^j \to M^j) \to \mathcal P (X^k \times X^k)$ is \\
		$(\vect{\sigma}, \vect{\sigma}') \in \Bf_{F_\mathcal N (\Gamma)} (\vect{x}, \cont)$ iff, for all $k$,
		\[ \sigma'_k \in \operatorname*{argmax}_{s \in X} g \left( s, (C^{T})\row{k} \vect{x} \oplus D\row{k} \cont \left(B^T \vect{x} \oplus D^T f(\vect{\sigma} \left[ k \mapsto s \right])\right) \oplus (E+E^T)\row{k} f(\vect{\sigma} \left[ k \mapsto s \right])\right) \]
	\end{itemize}
\end{theorem}
\begin{proof}
See Appendix~\ref{app:def_functor} on page~\pageref{proof:th_game_on_graph}.
\end{proof}

\section{Examples}~\label{sec:example}

We return to examples:
the majority (Example~\ref{ex:majority}), the best-shot public goods  (Example~\ref{ex:best-shot_public_goods}) and
the weakest-link public goods (Example~\ref{ex:weakest-link_public_goods}) games, and demonstrate various applications of our framework.
We first show that to compute the Nash equilibrium of the
majority game played on interconnected cliques is to calculate equilibria of its clique subgames.% Example~\ref{ex_f:majority} shows a case where a compositional analysis is possible. In other cases, as in Example~\ref{ex_f:weakest-link_public_goods}, where the compositional analysis is not possible but compositionality still gives conceptual clarity in the modelling of the game.
\begin{example}[Majority game]~\label{ex_f:majority}
  In the majority game the best response can be decomposed into the
  best responses of its components. Let $\N$ be the monoid
  network game for the majority game, defined on
  pg.~\pageref{ex_m:majority},
    and consider a graph composed of $N$
  cliques, as follows:
  \begin{itemize}
    \item each vertex of each clique can be connected to at most one
      vertex of another clique,
    \item in each clique there is at least one vertex not
      connected to any vertex outside its clique.
  \end{itemize} Such graphs decompose as $N$ open graphs,
   each a clique with some boundary connections.
  We omit the details and give, instead, an illustrative example:
  below left is a picture of three connected cliques, while the schematic on the right is the corresponding expression in $\Graph$.
  \\
  \graphmajority

  It is easy to show
  % (Appendix~\ref{app:examples}, Proposition~\ref{prop:B_majority})
  that the choice
  of each clique does not depend on the choices of other
  cliques.
  Indeed, the Nash equilibria of the majority game
  played on connected cliques in our sense are those strategy profiles where, in every clique, all players make the same choice. In
  particular, there are $2^N$ Nash equilibria.

  \vspace{.1cm}
  \begin{minipage}{0.65 \textwidth}
  In some cases, players can take into account the choice of
  another player with a different intensity. This can be modelled
  by changing the number of edges between the vertices. Let us consider the above example with some of the vertices connected multiple times.
  This modification of the network---illustrated on the right---reflects in a
  modification of the equilibria, which are now strategy
  profiles in which every player takes the same choice.
  \end{minipage}
  \begin{minipage}{0.35 \textwidth}
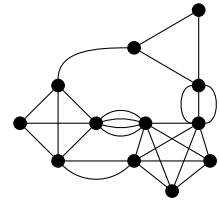
  \begin{center} \multigraphmajority
  \end{center}
  \end{minipage}
\end{example}

In the best-shot public goods game (Example~\ref{ex:best-shot_public_goods}), the Nash equilibrium is when no player
invests. In Example~\ref{ex_f:best-shot_public_goods}, we show how the compositional description is useful to adapt the model to a slightly different situation. We can imagine that one of the players now has access to incentives to invest in the public good. This scenario is represented by modifying the game and allowing one player to interact with the environment, which is the source of the incentives for this player.
This modification ``opens'' the game to one of type
$\diset{1}{1} \to \diset{X}{\R}$: as a result, the Nash
equilibrium changes.
This is a simple model of a common economic situation, `solving' a social dilemma by external intervention, for example by regulation \cite{golub_etal_targeting_interventions}.

\begin{example}[Best-shot public goods game]~\label{ex_f:best-shot_public_goods}
  Consider the best-shot public goods game played on a graph that
  contains a vertex connected to all other vertices.
  Removing the central vertex from this graph leaves an open
  graph that we will call $\Gamma$.
  \\
  \graphbestshot

  Here, $F_\N(\Gamma)$ is the best-shot public goods game played on
  the open graph $\Gamma$, $p$ is the central player that has
  been substituted, and $S$ is
  the external open game that influences $p$. The utility function
  of player $p$ and the coplay function of $S$ are as follows.
  \\
  \begin{minipage}{.6 \textwidth}
    \[u_{p}(\Gamma,x) = \begin{cases}
      1-c+\delta & \text{if } x_p = 1\\ 1-\epsilon & \text{if
      } x_p = 0 \land \exists (p,j) \in E_{\Gamma} \ x_j=1\\ 1-c & \text{if
      } \forall j \ x_j=0
    \end{cases}\]
  \end{minipage}
  \begin{minipage}{.4 \textwidth}
    \begin{flushright}
      \begin{align*} &\game{\textit{S}}{X}{\R}{1}{1}{1} \\ &\Cf(*,x,*)= \begin{cases} \delta & \text{if } x=1\\
          -\epsilon &\text{if } x=0
        \end{cases}
      \end{align*}
    \end{flushright}
  \end{minipage} \\
  The addition of the open game $S$ and the modification of player $p$ modifies the
  Nash equilibrium to be the strategy profile where only the central
  player invests. The idea is that the
  ``external'' agent $S$ incentivises the central player $p$ to
  invest. %The details are in Appendix~\ref{app:examples}.
\end{example}

Our last example illustrates a common situation where the compositional description of a game does \emph{not} allow a compositional analysis of the best response.
However, in this case, compositionality can be used to obtain a variant of the weakest-link public goods game  (Example~\ref{ex:weakest-link_public_goods}) where different cost
functions are used in different parts of the graph $G$.
The desired game is obtained by composing such open
games according to the structure of $G$.

\begin{example}[Weakest-link public goods game]~\label{ex_f:weakest-link_public_goods} Consider the
  weakest-link public goods game played on a connected graph
  $G$. Suppose that players have different cost functions.
  We partition them according to their cost functions, and use this partition to decompose the $G$ into an expression in $\Graph$, as illustrated for a particular example below: %\todo{Are the numbers of wires correct? Maybe go for a simpler topology?}
   \\
  \graphweakestlink

  While the definition above uses our compositional
  techniques, the Nash equilibrium %(Appendix~\ref{app:examples}, Propositions~\ref{prop:CN_B_weakest-link} and~\ref{prop:B_weakest-link})
  is calculated on the resulting closed game, and is a strategy profile where every player invests equally, with utility depending on individual cost
  functions.
  While it may be unsatisfying, this failure of Nash equilibria to be compositional can be seen as an inherent feature of game theory. In particular it is already present in the theory of open games; the passage from graphs to games is nevertheless fully compositional.
\end{example}

\section{Conclusions}

Our contribution is a compositional account of network games via strict monoidal functors. This adds a class of network games to the games that have been expressed in compositional game theory~\cite{hedges_etal_compositional_game_theory,bayesian_open_games}.
Of independent interest is our work on the category $\Graph$, extending~\cite{chantawibul_sobocinski_towards_compositional_graph_theory}. This is an approach to ``open graphs'' that, as we have seen, is compatible with the structure of open games, and in future work we will identify other uses of this category.

%It can also be seen as an instance of a
%functorial semantics of the category of open graphs $\Graph$, where the
%semantics of an open graph is given by the open game played on
%it.

%We defined a strict monoidal functor $\F \colon \Graph \to \Game$ for
%every monoid network game $\N$ and gave an explicit presentation of it
%(Theorem~\ref{th:game_on_graph}). In order to do this, we gave a
%combinatorial presentation of the category of open multigraphs
%$\Graph$.

We also intend to
extend the class of open graphs to \emph{directed} open graphs.
%This would require the study of a
%category of directed open graphs.
The motivation for this is that, in some network games, interactions between players are \emph{not} bidirectional. Consider, for example,
a variant of the majority game where there is an ``influencer'': a player whose choice affects the choices of other players, but is not in turn conversely affected.

We will also extend the menagerie of games that can be
played on a graph. We plan to study games with more
generic utility functions, incomplete information, and repeated games.
%On the game theory side,
It could also prove interesting to
study natural transformations between the functors that
define games, and explore the game theoretical relevance of such transformations.
% that they
% might have as morphisms of open games.

%\subsection*{Acknowledgements}
%Elena Di Lavore and Pawe\l \ Soboci\'nski were supported by the
%European Union through the ESF funded Estonian IT Academy
%research measure (project 2014-2020.4.05.19-0001).

\bibliography{ms}
%\newpage
\appendix
\section{Proofs for Section~\ref{sec:graphs}}~\label{app:graphs}

\begin{proof}[Details for definition~\ref{def:A}]\label{details_equivalence_A}
  By naturality of the symmetries, the vertex generators commute
  with any permutation matrix $P$: \vertexcommutepermutation.
  
  Thus, we can show that
  $\Gamma = (k, \left[ A \right] , B, C, D, \left[ E \right])$ and
  $\Gamma' = (k, \left[ A \right], B, CP^T, PD, \left[ PEP^T \right])$ represent
  the same open graph.
  \begin{center}
    \equivalentAequalGraph
  \end{center}
\end{proof}
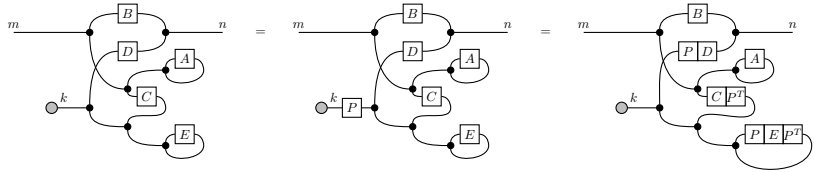
\begin{lemma}\label{lem:aprop}
$\A$ is a prop.
\end{lemma}
\begin{proof}
  We start by proving that $\A$ is a category. The diagram below can be rewritten, using the axioms of \(\bialg\), as a diagram of the form shown in Definition \ref{def:A}. The components of the normal form obtained in this way give the algebraic definition of the composition.
  % on the left, which is the graphical composition of the two morphisms, can be rewritten, using the axioms of $\bialg$, as the diagram on the right, which represents the definition of composition given as a formula.
\begin{equation*}
\Gamma' \circ \Gamma = \left(k+k', \left[ A + B A' B^T \right], BB', \left(C + B(A'+A'^T)D^T \vert BC' \right), \left( \begin{smallmatrix} DB'\\ D' \end{smallmatrix} \right) , \left[ \left( \begin{smallmatrix} E+DA'D^T & DC' \\ \mathbb{0} & E' \end{smallmatrix} \right) \right] \right)
\end{equation*}
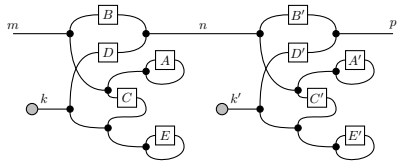
\begin{center}
  \compositionpropAdef
\end{center}

Identities are defined in the obvious way:
\(\mathbb{1}_n = (0, \left[ \mathbb{0}_n \right], \mathbb{1}_n, !, \text{!`}, \left[ ( \ ) \right] )\).

The definition of composition is coherent with the equivalence classes because, whenever \(\Gamma \sim \Gamma_{0}\) with matrix \(P\) and \(\Gamma' \sim \Gamma'_{0}\) with matrix \(P'\), \(\Gamma' \circ \Gamma \sim \Gamma'_{0} \circ \Gamma_{0}\) with matrix \(\left( \begin{smallmatrix} P & \mathbb{0} \\ \mathbb{0} & P' \end{smallmatrix} \right)\).
\removed{We need to check that the definition of composition is coherent with the equivalence classes.
Let $\Gamma = (k, \left[ A \right] , B, C, D, \left[ E \right]) \sim (k, \left[ A \right] , B, CP^T, PD, \left[ PEP^T \right]) = \Gamma_0 $ and $\Gamma' = (k', \left[ A' \right] , B', C', D', \left[ E' \right]) \sim (k', \left[ A' \right] , B', C'P'^T, P'D', \left[ P'E'P'^T \right]) = \Gamma'_0 $.\\
\begin{align*}
& \Gamma'_0 \circ \Gamma_0 \\
=& \begin{multlined}[t]
( k+k', \left[ A+BA'B^T \right], BB', (CP^T+B(A'+A'^T)D^TP^T \vert BC'P'^T),\\
\left( \begin{smallmatrix} PDB'\\ P'D' \end{smallmatrix} \right) , \left[ \left( \begin{smallmatrix} PEP^T+PDA'D^TP^T & PDC'P'^T \\ \mathbb{0} & P'E'P'^T \end{smallmatrix} \right) \right] )
\end{multlined}\\
\sim & \Gamma' \circ \Gamma
\end{align*}

with permutation matrix $\left( \begin{smallmatrix} P & \mathbb{0} \\ \mathbb{0} & P' \end{smallmatrix} \right)$.\\}
Composition is associative because the matrices relative to the
vertices are $[\ ]$-equivalent.
\removed{\begin{align*}
&\Gamma'' \circ (\Gamma' \circ \Gamma)\\
=& \begin{multlined}[t] (k+k'+k'', \left[ A + B(A'+B'A''B'^T) B^T \right], BB'B'', \\
(C + B(A'+A'^T+B'(A''+A''^T)B'^T))D^T \vert B(C'+B'(A''+A''^T)D'^T \vert B(B'C'')), \\
\left( \begin{smallmatrix} DB'B''\\ D'B'' \\ D'' \end{smallmatrix} \right) , \left[ \left( \begin{smallmatrix} E+D(A'+B'A''B'^T)D^T & D(C'+B'A''D'^T) & D(B'C'') \\ D'A''B'^T D^T & E'+D'A''D'^T & D'C'' \\ \mathbb{0} & \mathbb{0} & E'' \end{smallmatrix} \right) \right] )
\end{multlined} \\
=& \begin{multlined}[t] (k+k'+k'', \left[ A + B(A'+B'A''B'^T) B^T \right], BB'B'', \\
(C + B(A'+A'^T+B'(A''+A''^T)B'^T))D^T \vert B(C'+B'(A''+A''^T)D'^T \vert B(B'C'')), \\
\left( \begin{smallmatrix} DB'B''\\ D'B'' \\ D'' \end{smallmatrix} \right) , \left[ \left( \begin{smallmatrix} E+D(A'+B'A''B'^T)D^T & D(C'+B'(A''+A''^T)D'^T) & D(B'C'') \\ \mathbb{0} & E'+D'A''D'^T & D'C'' \\ \mathbb{0} & \mathbb{0} & E'' \end{smallmatrix} \right) \right] )
\end{multlined}\\
=& (\Gamma'' \circ \Gamma') \circ \Gamma
\end{align*}}
Clearly, composition is unital and we proved that $\A$ is a category. Now we prove that it is monoidal.\\
Lead by the interpretation of the matrices that define a morphism, we define monoidal product as follows.
\begin{equation*}
\Gamma \otimes \Gamma' = \left( k+k', \left[ \left( \begin{smallmatrix} A & \mathbb{0} \\ \mathbb{0} & A' \end{smallmatrix} \right) \right], \left( \begin{smallmatrix} B & \mathbb{0} \\ \mathbb{0} & B' \end{smallmatrix} \right), \left( \begin{smallmatrix} C & \mathbb{0} \\ \mathbb{0} & C' \end{smallmatrix} \right), \left( \begin{smallmatrix} D & \mathbb{0} \\ \mathbb{0} & D' \end{smallmatrix} \right), \left[ \left( \begin{smallmatrix} E & \mathbb{0} \\ \mathbb{0} & E' \end{smallmatrix} \right) \right] \right)
\end{equation*}
The monoidal unit is the empty diagram:
\(\mathbb{I} = (0, \left[ ( \ ) \right], ( \ ), ( \ ), ( \ ), \left[ ( \ ) \right] )\)

The monoidal product is well-defined on equivalence classes because, whenever \(\Gamma \sim \Gamma_{0}\) with matrix \(P\) and \(\Gamma' \sim \Gamma'_{0}\) with matrix \(P'\), \(\Gamma \otimes \Gamma' \sim \Gamma_{0} \otimes \Gamma'_{0}\) with matrix \(\left( \begin{smallmatrix} P & \mathbb{0} \\ \mathbb{0} & P' \end{smallmatrix} \right)\).
\removed{We verify that monoidal product is well-defined on equivalence classes. Let
\begin{align*}
\Gamma = (k, \left[ A \right] , B, C, D, \left[ E \right]) \sim & (k, \left[ A \right] , B, CP^T, PD, \left[ PEP^T \right]) = \Gamma_0\\
 \Gamma' = (k', \left[ A' \right] , B', C', D', \left[ E' \right]) \sim & (k', \left[ A' \right] , B', C'P'^T, P'D', \left[ P'E'P'^T \right]) = \Gamma'_0
\end{align*}
Then,
\begin{align*}
& \Gamma_0 \otimes \Gamma'_0 \\
&= \left( k+k', \left[ \left( \begin{smallmatrix} A & \mathbb{0} \\ \mathbb{0} & A' \end{smallmatrix} \right) \right], \left( \begin{smallmatrix} B & \mathbb{0} \\ \mathbb{0} & B' \end{smallmatrix} \right), \left( \begin{smallmatrix} CP^T & \mathbb{0} \\ \mathbb{0} & C'P'^T \end{smallmatrix} \right), \left( \begin{smallmatrix} PD & \mathbb{0} \\ \mathbb{0} & P'D' \end{smallmatrix} \right), \left[ \left( \begin{smallmatrix} PEP^T & \mathbb{0} \\ \mathbb{0} & P'E'P'^T \end{smallmatrix} \right) \right] \right)\\
& \sim \Gamma \otimes \Gamma'
\end{align*}
with permutation matrix $\left( \begin{smallmatrix} P & \mathbb{0} \\ \mathbb{0} & P' \end{smallmatrix} \right)$, which is the monoidal product of $P$ and $P'$.\\}
Clearly, monoidal product is strictly associative and unital.
Therefore, the pentagon and the triangle equations~\cite{macLane} hold trivially. The monoidal product is a bifunctor because \((\Gamma_0 \circ \Gamma) \otimes (\Gamma_0' \circ \Gamma_0) \sim  (\Gamma_0 \otimes \Gamma'_0) \circ (\Gamma \otimes \Gamma')\)
\removed{\begin{align*}
&(\Gamma_0 \circ \Gamma) \otimes (\Gamma_0' \circ \Gamma_0) \\
=& \begin{split}
	(k+k_0+k'+k_0', & \left[ \left(
		\begin{smallmatrix}
			A+BA_0A^T & \mathbb{0} \\
			\mathbb{0} & A'+B'A'_0A'^T
		\end{smallmatrix} \right) \right], \left(
		\begin{smallmatrix}
			BB_0 & \mathbb{0} \\
			\mathbb{0} & B'B'_0
		\end{smallmatrix} \right),\\
	&\left(
		\begin{smallmatrix}
			C+B(A_0+A_0^T)D^T & BC_0 & \mathbb{0} & \mathbb{0} \\
			\mathbb{0} & \mathbb{0} & C'+B'(A'_0+A'^T_0)D'^T  & B'C'_0
		\end{smallmatrix} \right),\\
	& \left(
		\begin{smallmatrix}
			DB_0 & \mathbb{0} \\
			D_0 & \mathbb{0} \\
			\mathbb{0} & D'B'_0 \\
			\mathbb{0} & D'_0
		\end{smallmatrix} \right), \left[ \left(
		\begin{smallmatrix}
			E+DA_0D^T & DC_0 & \mathbb{0} & \mathbb{0} \\
			\mathbb{0} & E_0 & \mathbb{0} & \mathbb{0} \\
			\mathbb{0} & \mathbb{0} & E'+D'A'_0D'^T & D'C'_0 \\
			\mathbb{0} & \mathbb{0} & \mathbb{0} & E'_0
		\end{smallmatrix} \right) \right] )
\end{split} \\
\sim & \begin{split}
	(k+k'+k_0+k_0', & \left[ \left(
		\begin{smallmatrix}
			A+BA_0A^T & \mathbb{0} \\
			\mathbb{0} & A'+B'A'_0A'^T
		\end{smallmatrix} \right) \right], \left(
		\begin{smallmatrix}
			BB_0 & \mathbb{0} \\
			\mathbb{0} & B'B'_0
		\end{smallmatrix} \right), \\
	& \left(
		\begin{smallmatrix}
			C+B(A_0+A_0^T)D^T & \mathbb{0} & BC_0 & \mathbb{0} \\
			\mathbb{0} & C'+B'(A'_0+A'^T_0)D'^T & \mathbb{0} & B'C'_0
		\end{smallmatrix} \right) , \\
	& \left(
		\begin{smallmatrix}
			DB_0 & \mathbb{0} \\
			\mathbb{0} & D'B'_0 \\
			D_0 & \mathbb{0} \\
			\mathbb{0} & D'_0
		\end{smallmatrix} \right), \left[ \left(
		\begin{smallmatrix}
			E+DA_0D^T & \mathbb{0} & DC_0 & \mathbb{0} \\
			\mathbb{0} & E'+D'A'_0D'^T & \mathbb{0} & D'C'_0 \\
			\mathbb{0} & \mathbb{0} & E_0 & \mathbb{0} \\
			\mathbb{0} & \mathbb{0} & \mathbb{0} & E'_0
		\end{smallmatrix} \right) \right] )
\end{split}\\
=& (\Gamma_0 \otimes \Gamma'_0) \circ (\Gamma \otimes \Gamma')
\end{align*}}
with permutation matrix $P = \left( \begin{smallmatrix} \mathbb{1} & \mathbb{0} & \mathbb{0} & \mathbb{0} \\  \mathbb{0} & \mathbb{0} & \mathbb{1} & \mathbb{0} \\  \mathbb{0} & \mathbb{1} & \mathbb{0} & \mathbb{0} \\ \mathbb{0} & \mathbb{0} & \mathbb{0} & \mathbb{1} \end{smallmatrix} \right)$.\\
Thus, $\A$ is a monoidal category. Finally, we prove that it is symmetric. Let $\sigma_{m,n}$ indicate the symmetry:
\(\sigma_{m,n} = (0, \left[ \mathbb{0} \right], \left( \begin{smallmatrix} \mathbb{0} & \mathbb{1}_m \\ \mathbb{1}_n & \mathbb{0} \end{smallmatrix} \right), !, \text{!`}, \left[ ( \ ) \right] )\)

Clearly, the symmetry is its own inverse: $ \sigma_{m,n} \circ \sigma_{n,m} = \mathbb{1}_{m+n}$.\\
Moreover, $\sigma$ is natural as \(\sigma_{n,n'} \circ (\Gamma \otimes \Gamma') \sim (\Gamma' \otimes \Gamma ) \circ \sigma_{m,m'}\)
\removed{\begin{align*}
\sigma_{n,n'} \circ (\Gamma \otimes \Gamma') =& (k+k', \left[ \left( \begin{smallmatrix} A & \mathbb{0} \\ \mathbb{0} & A' \end{smallmatrix} \right) \right], \left( \begin{smallmatrix} \mathbb{0} & B \\ B' & \mathbb{0} \end{smallmatrix} \right), \left( \begin{smallmatrix} C & \mathbb{0} \\ \mathbb{0} & C' \end{smallmatrix} \right), \left( \begin{smallmatrix} \mathbb{0} & D \\ D' & \mathbb{0} \end{smallmatrix} \right), \left[ \left( \begin{smallmatrix} E & \mathbb{0} \\ \mathbb{0} & E' \end{smallmatrix} \right) \right] ) \\
\sim & (k+k', \left[ \left( \begin{smallmatrix} A & \mathbb{0} \\ \mathbb{0} & A' \end{smallmatrix} \right) \right], \left( \begin{smallmatrix} \mathbb{0} & B \\ B' & \mathbb{0} \end{smallmatrix} \right), \left( \begin{smallmatrix} \mathbb{0} & C \\ C' & \mathbb{0} \end{smallmatrix} \right), \left( \begin{smallmatrix} D' & \mathbb{0} \\ \mathbb{0} & D \end{smallmatrix} \right), \left[ \left( \begin{smallmatrix} E' & \mathbb{0} \\ \mathbb{0} & E \end{smallmatrix} \right) \right] ) \\
= & (\Gamma' \otimes \Gamma ) \circ \sigma_{m,m'}
\end{align*}}
with permutation matrix $P = \left( \begin{smallmatrix} \mathbb{0} & \mathbb{1} \\ \mathbb{1} & \mathbb{0} \end{smallmatrix} \right) $.\\
Lastly, the symmetry satisfies the hexagon equations.
\removed{\begin{align*}
(\mathbb{1}_n \otimes \sigma_{p,m}) \circ \sigma_{m,n+p} =& (0, \left[ \mathbb{0} \right], \left( \begin{smallmatrix} \mathbb{0} & \mathbb{0} & \mathbb{1}_m \\ \mathbb{1}_n & \mathbb{0} & \mathbb{0} \\ \mathbb{0} & \mathbb{1}_p & \mathbb{0} \end{smallmatrix} \right) \left( \begin{smallmatrix} \mathbb{1}_n & \mathbb{0} & \mathbb{0} \\ \mathbb{0} & \mathbb{0} & \mathbb{1}_p \\ \mathbb{0} & \mathbb{1}_m & \mathbb{0} \end{smallmatrix} \right), !, \text{!`}, \left[ ( \ ) \right] ) \\
=& (0, \left[ \mathbb{0} \right], \left( \begin{smallmatrix} \mathbb{0} & \mathbb{1}_m & \mathbb{0} \\ \mathbb{1}_n & \mathbb{0} & \mathbb{0} \\ \mathbb{0} & \mathbb{0} & \mathbb{1}_p \end{smallmatrix} \right), !, \text{!`}, \left[ ( \ ) \right] )\\
= & \sigma_{m,n} \otimes \mathbb{1}_p
\end{align*}
\begin{align*}
(\sigma_{p,m} \otimes \mathbb{1}_n) \circ \sigma_{m+n,p} =& (0, \left[ \mathbb{0} \right], \left( \begin{smallmatrix} \mathbb{0} & \mathbb{1}_m & \mathbb{0} \\ \mathbb{0} & \mathbb{0} & \mathbb{1}_n \\ \mathbb{1}_p & \mathbb{0} & \mathbb{0} \end{smallmatrix} \right) \left( \begin{smallmatrix} \mathbb{0} &\mathbb{1}_p & \mathbb{0} \\ \mathbb{1}_m & \mathbb{0} & \mathbb{0} \\ \mathbb{0} & \mathbb{0} & \mathbb{1}_n \end{smallmatrix} \right), !, \text{!`}, \left[ ( \ ) \right] ) \\
=& (0, \left[ \mathbb{0} \right], \left( \begin{smallmatrix} \mathbb{1}_m & \mathbb{0} & \mathbb{0} \\ \mathbb{0} & \mathbb{0} & \mathbb{1}_n \\ \mathbb{0} & \mathbb{1}_p & \mathbb{0} \end{smallmatrix} \right), !, \text{!`}, \left[ ( \ ) \right] ) \\
=& \mathbb{1}_m \otimes \sigma_{n,p}
\end{align*}}
Thus, $\A$ is a symmetric monoidal category whose objects are natural numbers. In other words, it is a prop.
\end{proof}

\begin{lemma}\label{lem:bijprop}
$\bij$ is a prop.
\end{lemma}
\begin{proof}
The proof proceeds exactly as the previous one. We will use diagrammatic calculus of $\mathbf{Mat}$ for the permutation matrix of the morphisms in order to make the proofs more readable. We start by proving that $\bij$ is a category.
Composition is well-defined on equivalence classes by the monoidal structure of $\mathbf{Mat}$.
Let
$(k,P) \sim (k, \left( \begin{smallmatrix} \mathbb{1}_m & \mathbb{0} \\ \mathbb{0} & \sigma \end{smallmatrix} \right) P) = (k,P')$
and
$(l, Q) \sim (l, \left( \begin{smallmatrix} \mathbb{1}_{m+k} & \mathbb{0} \\ \mathbb{0} & \rho \end{smallmatrix} \right) Q) = (l,Q')$.
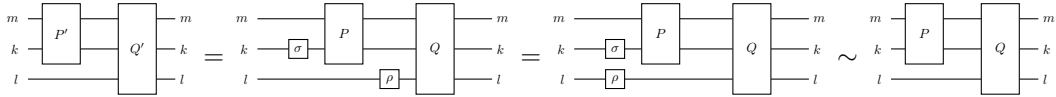
\[\compositioneqclassesbijFirst = \compositioneqclassesbijSecond = \compositioneqclassesbijThird \sim \compositioneqclassesbijFourth\]
with permutation matrix
$\left( \begin{smallmatrix} \mathbb{1}_m & \mathbb{0} & \mathbb{0} \\ \mathbb{0} & \sigma & \mathbb{0} \\ \mathbb{0} & \mathbb{0} & \rho \end{smallmatrix} \right)$.
Composition is clearly associative and unital because it is associative and unital in $\mathbf{Mat}$.
The monoidal product is defined with a symmetry on the left because we need to keep track of which of the inputs are bound.
\begin{equation*}
  (k,P) \otimes (k',P') = (k+k', \left( \begin{smallmatrix} \mathbb{1}_m & \mathbb{0} & \mathbb{0} & \mathbb{0} \\ \mathbb{0} & \mathbb{0} & \mathbb{1}_{m'} & \mathbb{0} \\ \mathbb{0} & \mathbb{1}_k & \mathbb{0} & \mathbb{0} \\ \mathbb{0} & \mathbb{0} & \mathbb{0} & \mathbb{1}_{k'} \end{smallmatrix} \right) \left( \begin{smallmatrix} P & \mathbb{0} \\ \mathbb{0} & P' \end{smallmatrix} \right) ) \qquad \defmonoidalbij
\end{equation*}
The monoidal unit is the empty diagram: \(\mathbb{I} = (0, ( \ ))\).
%\[ \defmonoidalunitbij\]
The monoidal product is well-defined on equivalence classes by naturality of the symmetries in $\mathbf{Mat}$. Let $(k,P) \sim (k, \left( \begin{smallmatrix} \mathbb{1}_m & \mathbb{0} \\ \mathbb{0} & \sigma \end{smallmatrix} \right) P) = (k,P')$ and $(l, Q) \sim (l, \left( \begin{smallmatrix} \mathbb{1}_n & \mathbb{0} \\ \mathbb{0} & \rho \end{smallmatrix} \right) Q) = (l,Q')$.
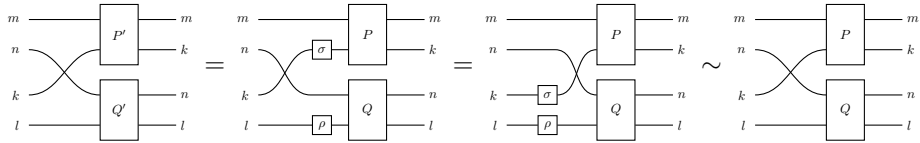
\[\monoidaleqclassesbijFirst = \monoidaleqclassesbijSecond = \monoidaleqclassesbijThird \sim \monoidaleqclassesbijFourth\]
with permutation matrix
$\left( \begin{smallmatrix} \mathbb{1}_{m+n} & \mathbb{0} & \mathbb{0} \\ \mathbb{0} & \sigma & \mathbb{0} \\ \mathbb{0} & \mathbb{0} & \rho \end{smallmatrix} \right)$.
The monoidal product is a bifunctor because we can change the order
in which we enumerate the vertices and because symmetries are
natural in $\mathbf{Mat}$.
\[\monoidalproductfunctorbijFirst \sim \monoidalproductfunctorbijSecond = \monoidalproductfunctorbijThird\]
with matrix \(\left( \begin{smallmatrix} \mathbb{1}_{m+m'+k} & \mathbb{0} & \mathbb{0} & \mathbb{0} \\ \mathbb{0} & \mathbb{0} & \mathbb{1}_{k'} & \mathbb{0} \\ \mathbb{0} & \mathbb{1}_l & \mathbb{0} & \mathbb{0} \\ \mathbb{0} & \mathbb{0} & \mathbb{0} & \mathbb{1}_{l'} \end{smallmatrix} \right)\).
The monoidal product is clearly unital. The symmetry is lifted from $\mathbf{Mat}$: \(\sigma_{m,n} = (0, \left( \begin{smallmatrix} \mathbb{0} & \mathbb{1}_m \\ \mathbb{1}_n & \mathbb{0} \end{smallmatrix} \right) )\).
\removed{\begin{flalign*}
  &(l,Q') \circ (k,P')&&\compositioneqclassesbijFirst\\
=& (k+l, \left( \begin{smallmatrix}
\left( \begin{smallmatrix} \mathbb{1}_m & \mathbb{0} \\
\mathbb{0} & \sigma \end{smallmatrix} \right) P & \mathbb{0} \\
\mathbb{0} & \mathbb{1}_l
\end{smallmatrix} \right) \left( \begin{smallmatrix}
\mathbb{1}_{m+k} & \mathbb{0} \\
\mathbb{0} & \rho \end{smallmatrix} \right) Q )&=&  \compositioneqclassesbijSecond\\
=& (k+l, \left( \begin{smallmatrix}
\mathbb{1}_m & \mathbb{0} & \mathbb{0} \\
\mathbb{0} & \sigma & \mathbb{0} \\
\mathbb{0} & \mathbb{0} & \rho \end{smallmatrix} \right) \left( \begin{smallmatrix}
P & \mathbb{0} \\
\mathbb{0} & \mathbb{1}_l \end{smallmatrix} \right) Q)& =&
\compositioneqclassesbijThird\\
\sim & (l,Q) \circ (k,P) & \sim & \compositioneqclassesbijFourth\\
\intertext{with permutation matrix
  $\left( \begin{smallmatrix} \mathbb{1}_m & \mathbb{0} & \mathbb{0} \\ \mathbb{0} & \sigma & \mathbb{0} \\ \mathbb{0} & \mathbb{0} & \rho \end{smallmatrix} \right)$.
  Composition is clearly associative because it is associative in $\mathbf{Mat}$.}
&\begin{aligned}
& ((l,R) \circ (k,Q)) \circ (j,P) \\
=& (j+k+l,  \left( \begin{smallmatrix} P & \mathbb{0} \\ \mathbb{0} & \mathbb{1}_{k+l} \end{smallmatrix} \right)  \left( \begin{smallmatrix} Q & \mathbb{0} \\ \mathbb{0} & \mathbb{1}_l \end{smallmatrix} \right) R) \\
=& (j+k+l, \left( \begin{smallmatrix} \left( \begin{smallmatrix} P & \mathbb{0} \\ \mathbb{0} & \mathbb{1}_{k} \end{smallmatrix} \right)Q & \mathbb{0} \\ \mathbb{0} & \mathbb{1}_l \end{smallmatrix} \right) R)\\
=& (l,R) \circ ((k,Q) \circ (j,P))
\end{aligned}&&\compositionassociativebij\\
\intertext{Composition is unital by unitality of the composition in $\mathbf{Mat}$.}
& (k,P) \circ (0,\mathbb{1}_m) = (0+k, \mathbb{1}_{m+k} P) = (k,P) && \compositionunitalbijFirst \\
& (0,\mathbb{1}_{m+k}) \circ (k,P) = (k+0, P \mathbb{1}_{m+k}) = (k,P) && \compositionunitalbijSecond\\
\intertext{The monoidal product is defined with a symmetry on the
  left because we need to keep track of which of the inputs are bound.}
  &(k,P) \otimes (k',P') = (k+k', \left( \begin{smallmatrix} \mathbb{1}_m & \mathbb{0} & \mathbb{0} & \mathbb{0} \\ \mathbb{0} & \mathbb{0} & \mathbb{1}_{m'} & \mathbb{0} \\ \mathbb{0} & \mathbb{1}_k & \mathbb{0} & \mathbb{0} \\ \mathbb{0} & \mathbb{0} & \mathbb{0} & \mathbb{1}_{k'} \end{smallmatrix} \right) \left( \begin{smallmatrix} P & \mathbb{0} \\ \mathbb{0} & P' \end{smallmatrix} \right) ) &&\defmonoidalbij\\
\intertext{The monoidal unit is the empty diagram.}
&\mathbb{I} = (0, ( \ )) &&\defmonoidalunitbij\\
\intertext{Monoidal product is well-defined on equivalence classes by naturality of the symmetries in $\mathbf{Mat}$. Let $(k,P) \sim (k, \left( \begin{smallmatrix} \mathbb{1}_m & \mathbb{0} \\ \mathbb{0} & \sigma \end{smallmatrix} \right) P) = (k,P')$ and $(l, Q) \sim (l, \left( \begin{smallmatrix} \mathbb{1}_n & \mathbb{0} \\ \mathbb{0} & \rho \end{smallmatrix} \right) Q) = (l,Q')$.}
&(k,P') \otimes (l,Q')& &\monoidaleqclassesbijFirst\\
=& (k+l, \left( \begin{smallmatrix} \mathbb{1}_m & \mathbb{0} & \mathbb{0} & \mathbb{0} \\ \mathbb{0} & \mathbb{0} & \mathbb{1}_{n} & \mathbb{0} \\ \mathbb{0} & \mathbb{1}_k & \mathbb{0} & \mathbb{0} \\ \mathbb{0} & \mathbb{0} & \mathbb{0} & \mathbb{1}_{l} \end{smallmatrix} \right) \left( \begin{smallmatrix} \left( \begin{smallmatrix} \mathbb{1}_m & \mathbb{0} \\ \mathbb{0} & \sigma \end{smallmatrix} \right) P & \mathbb{0} \\ \mathbb{0} & \left( \begin{smallmatrix} \mathbb{1}_n & \mathbb{0} \\ \mathbb{0} & \rho \end{smallmatrix} \right) Q \end{smallmatrix} \right) ) &=&\monoidaleqclassesbijSecond\\
=&(k+l, \left( \begin{smallmatrix} \mathbb{1}_m & \mathbb{0} & \mathbb{0} & \mathbb{0} \\ \mathbb{0} & \mathbb{1}_{n} & \mathbb{0} & \mathbb{0} \\ \mathbb{0} & \mathbb{0} & \sigma & \mathbb{0} \\ \mathbb{0} & \mathbb{0} & \mathbb{0} & \rho \end{smallmatrix} \right) \left( \begin{smallmatrix} \mathbb{1}_m & \mathbb{0} & \mathbb{0} & \mathbb{0} \\ \mathbb{0} & \mathbb{0} & \mathbb{1}_{n} & \mathbb{0} \\ \mathbb{0} & \mathbb{1}_k & \mathbb{0} & \mathbb{0} \\ \mathbb{0} & \mathbb{0} & \mathbb{0} & \mathbb{1}_{l} \end{smallmatrix} \right) \left( \begin{smallmatrix} P & \mathbb{0} \\ \mathbb{0} & Q \end{smallmatrix} \right) ) &=&\monoidaleqclassesbijThird\\
\sim& (k,P) \otimes (l,Q) &\sim &\monoidaleqclassesbijFourth\\
\intertext{with permutation matrix
  $\left( \begin{smallmatrix} \mathbb{1}_{m+n} & \mathbb{0} & \mathbb{0} \\ \mathbb{0} & \sigma & \mathbb{0} \\ \mathbb{0} & \mathbb{0} & \rho \end{smallmatrix} \right)$.
  Monoidal product is a functor because we can change the order
  in which we enumerate the vertices and because symmetries are
  natural in $\mathbf{Mat}$. For the calculations below, we
  indicate with
  $C =\left( \begin{smallmatrix} \mathbb{1}_m & \mathbb{0} & \mathbb{0} & \mathbb{0} \\ \mathbb{0} & \mathbb{0} & \mathbb{1}_{m'} & \mathbb{0} \\ \mathbb{0} & \mathbb{1}_{k+l} & \mathbb{0} & \mathbb{0} \\ \mathbb{0} & \mathbb{0} & \mathbb{0} & \mathbb{1}_{k'+l'} \end{smallmatrix} \right) \left( \begin{smallmatrix}  \left( \begin{smallmatrix} P & \mathbb{0} \\ \mathbb{0} & \mathbb{1}_l \end{smallmatrix} \right) Q & \mathbb{0} \\ \mathbb{0} &  \left( \begin{smallmatrix} P' & \mathbb{0} \\ \mathbb{0} & \mathbb{1}_{l'} \end{smallmatrix} \right) Q' \end{smallmatrix} \right)$
  the matrix associated to
  $((l,Q) \circ (k,P)) \otimes ((l',Q') \circ (k',P'))$.}
&((l,Q) \circ (k,P)) \otimes ((l',Q') \circ (k',P'))& & \\
=& (k+l+k'+l', C ) &&\monoidalproductfunctorbijFirst\\
\sim& (k+k'+l+l', \left( \begin{smallmatrix} \mathbb{1}_{m+m'+k} & \mathbb{0} & \mathbb{0} & \mathbb{0} \\ \mathbb{0} & \mathbb{0} & \mathbb{1}_{k'} & \mathbb{0} \\ \mathbb{0} & \mathbb{1}_l & \mathbb{0} & \mathbb{0} \\ \mathbb{0} & \mathbb{0} & \mathbb{0} & \mathbb{1}_{l'} \end{smallmatrix} \right) C )
&\sim &\monoidalproductfunctorbijSecond\\
=& ((l,Q) \otimes (l',Q')) \circ ((k,P) \otimes (k',P')) &=&\monoidalproductfunctorbijThird\\
\intertext{Monoidal product is clearly unital. The symmetry is lifted from $\mathbf{Mat}$.}
& \sigma_{m,n} = (0, \left( \begin{smallmatrix} \mathbb{0} & \mathbb{1}_m \\ \mathbb{1}_n & \mathbb{0} \end{smallmatrix} \right) ) &&\swapmn\\
\end{flalign*}}
The symmetry is its own inverse and it satisfies the hexagon equations because it does so in $\mathbf{Mat}$.\\
Therefore, $\bij$ is a prop.\\
\end{proof}

\begin{lemma}\label{lem:bij_is_free_full}
$\bij$ is isomorphic to the free prop on one generator $0 \to 1$.
\end{lemma}
\begin{proof}
Define \(\alpha^{\#} \colon \bij \longrightarrow \mathbb{P}\) to be \(\alpha^{\#}(k,P) = P \circ (\mathbb{1}_m \otimes \bigotimes_k v)\),
% \begin{align*}
% \alpha^{\#} \colon \bij & \longrightarrow \mathbb{P} &
% \alpha^{\#} \colon (k,P) & \longmapsto P \circ (\mathbb{1}_m \otimes \bigotimes_k v)\\
% \deffreemapbijLHS&\longmapsto\deffreemapbijRHS
% \end{align*}
where $P \in \mathbb{P}$ is the product of the symmetries that form $P$ in $\bij$. Diagrammatically, \(\freemapbijeqclassesFifth = \deffreemapbijRHS\). \\
We prove that $\alpha^{\#}$ is well-defined on equivalence classes. Let $(k,P) \sim (k, \left( \begin{smallmatrix} \mathbb{1}_m & \mathbb{0} \\ \mathbb{0} & \sigma \end{smallmatrix} \right) P) = (k,P')$.
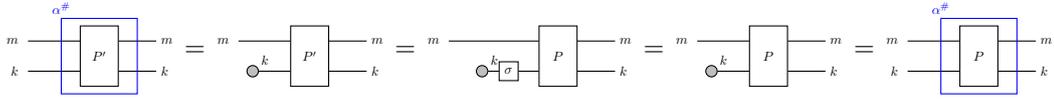
\[\freemapbijeqclassesFirst = \freemapbijeqclassesSecond = \freemapbijeqclassesThird = \freemapbijeqclassesFourth = \freemapbijeqclassesFifth\]
\removed{\begin{flalign*}
&\alpha^{\#}(k,P')&&\freemapbijeqclassesFirst\\
=& \left( \left( \begin{smallmatrix} \mathbb{1}_m & \mathbb{0} \\ \mathbb{0} & \sigma \end{smallmatrix} \right) P \right) \circ (\mathbb{1}_m \otimes v^k)&=&\freemapbijeqclassesSecond\\
=& P \circ (\mathbb{1}_m \otimes \sigma) \circ (\mathbb{1}_m \otimes v^k)&=&\freemapbijeqclassesThird\\
=& P \circ (\mathbb{1}_m \otimes \sigma \circ v^k)&&\\
=& P \circ (\mathbb{1}_m \otimes v^k)&=&\freemapbijeqclassesFourth\\
=& \alpha^{\#}(k,P)&=&\freemapbijeqclassesFifth\\
\end{flalign*}}
We show graphically that $\alpha^{\#}$ is a prop homomorphism\\
\[\alpha^{\#}(\mathbb{I}) = \qquad = \mathbb{I} \qquad
\alpha^{\#}(\mathbb{1}_n) = \identitywire = \mathbb{1}_n \qquad
\alpha^{\#}(0,\sigma) = \swapbox = \sigma\]
{\renewcommand{\arraystretch}{3}
\begin{tabular}{r c l}
$\alpha^{\#}((l,Q) \circ (k,P)) =$&\freemapbijcomposition&$= \alpha^{\#}(l,Q) \circ \alpha^{\#}(k,P)$\\
$\alpha^{\#}((k,P)\otimes (k',P')) =$&\freemapbijmonoidalproduct&$= \alpha^{\#}(k,P)\otimes \alpha^{\#}(k',P')$\\
\end{tabular}}\\

and, by its definition,
$$\alpha^{\#}(\phi) = v$$
Moreover, $\alpha^{\#}$ is the unique morphism $\bij \to \mathbb{P}$ with this property. In fact, suppose there is $\beta \colon \bij \to \mathbb{P}$ such that $\beta(\phi) = v$.
Then,
\begin{align*}
\beta (k,P) =& \beta((0,P) \circ ((0,\mathbb{1}_n) \otimes (k,\mathbb{1}_k)))
= \beta(0,P) \circ (\beta(0,\mathbb{1}_n) \otimes \beta(k,\mathbb{1}_k)) \\
=& P \circ (\mathbb{1}_n \otimes \bigotimes_k v)
= a^{\#} (k,P)
\end{align*}
Then $\bij$ is isomorphic to the free prop over one generator $ 0 \to 1 $.
\end{proof}

\begin{claim}\label{cl:injections-are-homs}
The following are prop homomorphisms.\\
\begin{minipage}{0.4 \textwidth}
    \begin{align*} i_1 \colon \Umat & \longrightarrow \A\\ n & \longmapsto n \\ (B, \left[ A \right] ) & \longmapsto (0, \left[ A \right], B, !, \text{!`}, \left[ ( \ ) \right] ) \\
       \definclusiononeALHS &\longmapsto \definclusiononeARHS
    \end{align*}
  \end{minipage}
  \begin{minipage}{0.5 \textwidth}
    \begin{align*} i_2 \colon \bij & \longrightarrow \A\\ n & \longmapsto n \\ (k,P) & \longmapsto (k, \left[ \mathbb{0}_n \right], P\row{[1,n]}, \mathbb{0}_{nk}, P\row{[n+1,n+k]}, \left[ \mathbb{0}_k \right] )\\
       \definclusiontwoALHS &\longmapsto \definclusiontwoARHS
    \end{align*}
  \end{minipage}
\end{claim}
\begin{proof}
  We prove graphically that they are prop homomorphisms.\\
  \[i_1(\mathbb{I}) = \qquad = \mathbb{I} \qquad
i_1(\mathbb{1}_n) = \identitywire = \mathbb{1}_n \qquad
i_1(\sigma, \left[ ( \ ) \right]) = \swapbox = \sigma\]
{\renewcommand{\arraystretch}{3}
\begin{tabular}{r c l}
$i_1((B', \left[ A' \right]) \circ (B, \left[ A \right])) =$ &\inclusiononeAcomposition&$= i_1(B', \left[ A' \right]) \circ i_1(B, \left[ A \right])$\\
$i_1((B, \left[ A \right]) \otimes (B', \left[ A' \right])) =$&
\inclusiononeAmonoidalproduct&$= i_1(B, \left[ A \right]) \otimes i_1(B', \left[ A' \right])$
\end{tabular}\\
\[i_2(\mathbb{I}) = \qquad = \mathbb{I} \qquad
i_2(\mathbb{1}_n) = \identitywire = \mathbb{1}_n \qquad
i_2(0,\sigma) = \swapbox = \sigma \]
\begin{tabular}{r c l}
$i_2((l,Q) \circ (k,P)) =$ &\freemapbijcomposition&$= i_2(l,Q) \circ i_1(k,P)$\\
$i_2((k,P)\otimes (k',P')) =$ &\freemapbijmonoidalproduct&$= i_2(k,P)\otimes i_2(k',P')$
\end{tabular}}\\
\end{proof}
\begin{lemma}\label{lem:Hhomo}
$H$, defined on page~\pageref{pg:Hdef}, is a prop homomorphism.
\end{lemma}
\begin{proof}
Recall that \(H \colon \A \to \mathcal{C}\) is identity on objects and \(H(k, \left[ A \right] , B, C, D, \left[ E \right]) = f_1 \left( \left( \begin{smallmatrix} B \\ D\end{smallmatrix} \right), \left[ \left( \begin{smallmatrix} A & C \\ \mathbb{0} & E \end{smallmatrix} \right) \right] \right) \circ \left( \mathbb{1}_m \otimes f_2(k, \mathbb{1}_k) \right)\).
% \begin{align*}
% H \colon \A &\longrightarrow \mathcal{C}\\
% n & \longmapsto n\\
% (k, \left[ A \right] , B, C, D, \left[ E \right]) & \longmapsto f_1 \left( \left( \begin{smallmatrix} B \\ D\end{smallmatrix} \right), \left[ \left( \begin{smallmatrix} A & C \\ \mathbb{0} & E \end{smallmatrix} \right) \right] \right) \circ \left( \mathbb{1}_m \otimes f_2(k, \mathbb{1}_k) \right) \\
% \defHLHS&\longmapsto\defHRHS
% \end{align*}
By calling $w = \left( \left( \begin{smallmatrix} B \\ D\end{smallmatrix} \right), \left[ \left( \begin{smallmatrix} A & C \\ \mathbb{0} & E \end{smallmatrix} \right) \right] \right)$, which is a morphism in $\Umat$, we can depict the image of \(H\) diagrammatically.
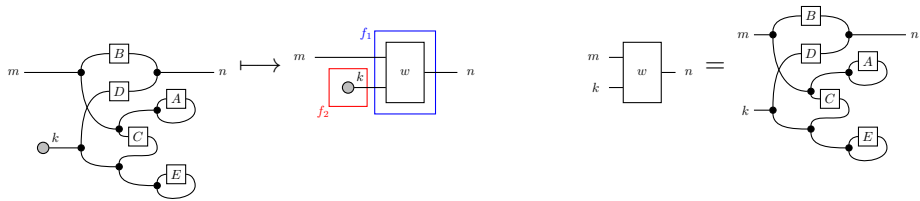
\begin{align*}
  \defHLHS & \longmapsto \defHRHS & \defw
\end{align*}

We need to prove that $H$ is well-defined on equivalence classes. Let $\Gamma = (k, \left[ A \right], B, C, D, \left[ E \right]) \sim (k, \left[ A \right], B, CP^T, PD, \left[ PEP^T \right]) = \Gamma'$.
\begin{align*}
  H(\Gamma') &= \HeqclassesFirst = \HeqclassesSecond = \HeqclassesThird \\
  &= \HeqclassesFourth = \HeqclassesFifth = H(\Gamma)
\end{align*}
We prove that $H$ is a prop homomorphism. Clearly, $H$ is
identity on objects. Moreover, it preserves composition, as it is
shown by the diagrams.
\begin{align*}
  H(\Gamma') \circ H(\Gamma) &= \HcompositionFirst = \HcompositionSecond \\
  &= \HcompositionThird = H(\Gamma' \circ \Gamma)
\end{align*}
$H$ preserves identities:
\(H( \mathbb{1}_n) = \Hidentity = \identitywire = \mathbb{1}_{n}\).

$H$ preserves monoidal product. This is also more clearly seen with string diagrams.
\begin{align*}
  H(\Gamma) \otimes H(\Gamma') &= \HmonoidalproductFirst = \HmonoidalproductSecond = \HmonoidalproductThird \\
  &= \HmonoidalproductFourth = H(\Gamma \otimes \Gamma')
\end{align*}
It is easy to show that $H$ preserves monoidal unit and symmetries.
\[H(\mathbb{I}) = \Hmonoidalunit = \nodiagram = \mathbb{I} \qquad H(\sigma_{m,n}) = \HswapFirst = \HswapSecond = \sigma_{m,n}\]
\removed{\begin{flalign*}
&H(\Gamma') = f_1 \left( \left( \begin{smallmatrix} B \\ PD\end{smallmatrix} \right), \left[ \left( \begin{smallmatrix} A & CP^T \\ \mathbb{0} & PEP^T \end{smallmatrix} \right) \right] \right) \circ \left( \mathbb{1}_m \otimes f_2(k, \mathbb{1}_k) \right)&&\HeqclassesFirst\\
=& f_1\left( \left( \left( \begin{smallmatrix} B \\ D\end{smallmatrix} \right), \left[ \left( \begin{smallmatrix} A & C \\ \mathbb{0} & E \end{smallmatrix} \right) \right] \right) \circ \left( \left( \begin{smallmatrix} \mathbb{1}_m & \mathbb{0} \\ \mathbb{0} & P \end{smallmatrix} \right), \left[ \mathbb{0} \right] \right) \right) \circ \left( \mathbb{1}_m \otimes f_2(k, \mathbb{1}_k) \right)&=&\HeqclassesSecond\\
=& f_1 \left( \left( \begin{smallmatrix} B \\ D \end{smallmatrix} \right), \left[ \left( \begin{smallmatrix} A & C \\ \mathbb{0} & E \end{smallmatrix} \right) \right] \right) \circ (\mathbb{1}_m \otimes P) \circ \left( \mathbb{1}_m \otimes f_2(k, \mathbb{1}_k) \right)&=&\HeqclassesThird\\
=& f_1 \left( \left( \begin{smallmatrix} B \\ D\end{smallmatrix} \right), \left[ \left( \begin{smallmatrix} A & C \\ \mathbb{0} & E \end{smallmatrix} \right) \right] \right) \circ \left( \mathbb{1}_m \otimes P \circ f_2(k, \mathbb{1}_k) \right)&=&\HeqclassesFourth\\
=& f_1 \left( \left( \begin{smallmatrix} B \\ D\end{smallmatrix} \right), \left[ \left( \begin{smallmatrix} A & C \\ \mathbb{0} & E \end{smallmatrix} \right) \right] \right) \circ \left( \mathbb{1}_m \otimes f_2(k, \mathbb{1}_k) \right) = H(\Gamma)&=&\HeqclassesFifth\\
\intertext{We prove that $H$ is a prop homomorphism. Clearly, $H$ is
identity on objects. Moreover, it preserves composition, as it is
shown by the diagrams.}
  &H(\Gamma') \circ H(\Gamma)&&\\
  =&\begin{aligned}[t]
       f_1 \left( \left( \begin{smallmatrix} B' \\ D' \end{smallmatrix} \right), \left[ \left( \begin{smallmatrix} A' & C' \\ \mathbb{0} & E' \end{smallmatrix} \right) \right] \right) \circ \left( \mathbb{1}_n \otimes f_2(k', \mathbb{1}_{k'}) \right) \\
      \circ f_1 \left( \left( \begin{smallmatrix} B \\ D\end{smallmatrix} \right), \left[ \left( \begin{smallmatrix} A & C \\ \mathbb{0} & E \end{smallmatrix} \right) \right] \right) \circ \left( \mathbb{1}_m \otimes f_2(k, \mathbb{1}_k) \right)
    \end{aligned}&&\HcompositionFirst\\
  =&\begin{aligned}[t]
       f_1 \left( \left( \begin{smallmatrix} B' \\ D' \end{smallmatrix} \right), \left[ \left( \begin{smallmatrix} A' & C' \\ \mathbb{0} & E' \end{smallmatrix} \right) \right] \right) \circ ( f_1 \left( \left( \begin{smallmatrix} B \\ D\end{smallmatrix} \right), \left[ \left( \begin{smallmatrix} A & C \\ \mathbb{0} & E \end{smallmatrix} \right) \right] \right) \otimes \mathbb{1}_k) \\
      \circ (\mathbb{1}_m \otimes f_2(k, \mathbb{1}_k) \otimes f_2(k', \mathbb{1}_{k'}) )
    \end{aligned}&=&\HcompositionSecond\\
  =&\begin{aligned}[t]
       f_1 \left( \left( \left( \begin{smallmatrix} B' \\ D' \end{smallmatrix} \right), \left[ \left( \begin{smallmatrix} A' & C' \\ \mathbb{0} & E' \end{smallmatrix} \right) \right] \right) \circ \left( \left( \left( \begin{smallmatrix} B \\ D\end{smallmatrix} \right), \left[ \left( \begin{smallmatrix} A & C \\ \mathbb{0} & E \end{smallmatrix} \right) \right] \right) \otimes \mathbb{1}_k \right) \right) \\
      \circ (\mathbb{1}_m \otimes f_2((k, \mathbb{1}_k) \otimes (k', \mathbb{1}_{k'})) )
    \end{aligned}&&\\
  =&\begin{aligned}[t]
       f_1 \left( \left( \begin{smallmatrix} BB' \\ DB' \\ D' \end{smallmatrix} \right), \left[ \left( \begin{smallmatrix} A+BA'B^T & C+BA'D^T & BC' \\ DA'B^T & E+DA'D^T & DC' \\ \mathbb{0} & \mathbb{0} & E' \end{smallmatrix} \right) \right] \right) \\
      \circ ( \mathbb{1}_m \otimes f_2(k+k', \mathbb{1}_{k+k'}))
    \end{aligned}&&\\
  =&\begin{aligned}[t]
       f_1 \left( \left( \begin{smallmatrix} BB' \\ DB' \\ D' \end{smallmatrix} \right), \left[ \left( \begin{smallmatrix} A+BA'B^T & C+B(A'+A'^T)D^T & BC' \\ \mathbb{0} & E+DA'D^T & DC' \\ \mathbb{0} & \mathbb{0} & E' \end{smallmatrix} \right) \right] \right) \\
      \circ ( \mathbb{1}_m \otimes f_2(k+k', \mathbb{1}_{k+k'}))
    \end{aligned}&=&\HcompositionThird\\
  =& H(\Gamma' \circ \Gamma) && \\
\intertext{$H$ preserves identities.}
&H( \mathbb{1}_n) = H(0, \left[ \mathbb{0}_n \right], \mathbb{1}_n, !, \text{!`}, \left[ ( \ ) \right] )&&\Hidentity\\
=&f_1(\mathbb{1}_n, \left[ \mathbb{0}_n \right]) \circ (\mathbb{1}_n \otimes f_2(0, \mathbb{1}_0)) = \mathbb{1}_n&=&\identitywire\\
\intertext{$H$ preserves monoidal product. This is also more clearly seen with the string diagrams.}
&H(\Gamma) \otimes H(\Gamma')&&\\
=&{\begin{aligned}[t]
& \left( f_1 \left( \left( \begin{smallmatrix} B \\ D\end{smallmatrix} \right), \left[ \left( \begin{smallmatrix} A & C \\ \mathbb{0} & E \end{smallmatrix} \right) \right] \right) \circ \left( \mathbb{1}_m \otimes f_2(k, \mathbb{1}_k) \right) \right) \\
& \otimes \left( f_1 \left( \left( \begin{smallmatrix} B' \\ D' \end{smallmatrix} \right), \left[ \left( \begin{smallmatrix} A' & C' \\ \mathbb{0} & E' \end{smallmatrix} \right) \right] \right) \circ \left( \mathbb{1}_n \otimes f_2(k', \mathbb{1}_{k'}) \right) \right)
\end{aligned}}&=&\HmonoidalproductFirst\\
=&{\begin{aligned}[t]
&  \left( f_1 \left( \left( \begin{smallmatrix} B \\ D\end{smallmatrix} \right), \left[ \left( \begin{smallmatrix} A & C \\ \mathbb{0} & E \end{smallmatrix} \right) \right] \right) \otimes f_1 \left( \left( \begin{smallmatrix} B' \\ D' \end{smallmatrix} \right), \left[ \left( \begin{smallmatrix} A' & C' \\ \mathbb{0} & E' \end{smallmatrix} \right) \right] \right) \right) \\
&\circ (\mathbb{1}_m \otimes \sigma_{m',k} \otimes \mathbb{1}_{k'}) \circ (\mathbb{1}_{m+m'} \otimes f_2(k, \mathbb{1}_k) \otimes f_2(k', \mathbb{1}_{k'}))
\end{aligned}}&=&\HmonoidalproductSecond\\
=&{\begin{aligned}[t]
& f_1 ( ( \left( \left( \begin{smallmatrix} B \\ D\end{smallmatrix} \right), \left[ \left( \begin{smallmatrix} A & C \\ \mathbb{0} & E \end{smallmatrix} \right) \right] \right) \otimes \left( \left( \begin{smallmatrix} B' \\ D' \end{smallmatrix} \right), \left[ \left( \begin{smallmatrix} A' & C' \\ \mathbb{0} & E' \end{smallmatrix} \right) \right] \right) )\\
& \circ \left( \left( \begin{smallmatrix} \mathbb{1}_m & \mathbb{0} & \mathbb{0} & \mathbb{0} \\ \mathbb{0} & \mathbb{0}  & \mathbb{1}_{m'} & \mathbb{0} \\ \mathbb{0} & \mathbb{1}_k & \mathbb{0} & \mathbb{0} \\ \mathbb{0} & \mathbb{0} & \mathbb{0} & \mathbb{1}_{k'} \end{smallmatrix} \right) , \left[ \mathbb{0}_{m+m'+k+k'} \right] \right) )\\
& \circ (\mathbb{1}_{m+m'} \otimes f_2(k+k', \mathbb{1}_{k+k'}))
\end{aligned}}&=&\HmonoidalproductThird\\
=&{\begin{aligned}[t]
& f_1 \left( \left( \begin{smallmatrix} B & \mathbb{0} \\ \mathbb{0} & B' \\ D & \mathbb{0} \\ \mathbb{0} & D' \end{smallmatrix} \right), \left[ \left( \begin{smallmatrix} A & \mathbb{0} & C & \mathbb{0} \\ \mathbb{0} & A' & \mathbb{0} & C' \\ \mathbb{0} & \mathbb{0} & E & \mathbb{0} \\ \mathbb{0} & \mathbb{0} & \mathbb{0} & E' \end{smallmatrix} \right) \right] \right)\\
& \circ (\mathbb{1}_{m+m'} \otimes f_2(k+k', \mathbb{1}_{k+k'}))
\end{aligned}}&=&\HmonoidalproductFourth\\
=& H(\Gamma \otimes \Gamma') &&\\
\intertext{It is easy to show that $H$ preserves monoidal unit.}
&H(\mathbb{I}) = H(0, \left[ ( \ ) \right], ( \ ), ( \ ), ( \ ), \left[ ( \ ) \right] )&&\Hmonoidalunit\\
=&f_1(( \ ), \left[ ( \ ) \right]) \circ (\mathbb{1}_0 \otimes f_2(0, \mathbb{1}_0)) = \mathbb{I}&=&\nodiagram\\
\intertext{Finally, $H$ preserves symmetries.}
&H(\sigma_{m,n}) = H(0, \left[ \mathbb{0}_{m+n} \right], \left( \begin{smallmatrix} \mathbb{0} & \mathbb{1}_m \\ \mathbb{1}_n & \mathbb{0} \end{smallmatrix} \right), ! , \text{!`}, \left[ ( \ ) \right])& &\\
=& f_1( \left( \begin{smallmatrix} \mathbb{0} & \mathbb{1}_m \\ \mathbb{1}_n & \mathbb{0} \end{smallmatrix} \right), \left[ \mathbb{0}_{m+n} \right]) \circ (\mathbb{1}_{m+n} \otimes f_2(0,\mathbb{1}_0))& &\HswapFirst\\
=& f_1(\sigma_{m,n}) = \sigma_{m,n}&=&\HswapSecond\\
\end{flalign*}}
\end{proof}

\section{Proofs for Sections~\ref{sec:def_functor}}~\label{app:def_functor}

\begin{proof}[Proof of Theorem~\ref{th:Fn_monoidal_functor}~\label{proof:th_F_functor}]
  $F_\N$ respects the equations of $\Graph$ because both the tuples
\newline $\left( \begin{tikzpicture}[baseline=-0.1cm,scale=.5,transform shape]
	\node [dot] (m2) at (0, 0) {};
		\draw [->-] (m2) [out=135, in=0] to (-1,0.5);
		\draw [->-] (m2) [out=-135, in=0] to (-1,-0.5);
		\draw [->-] (1, 0) to (m2);
\end{tikzpicture} , \begin{tikzpicture}[baseline=-0.1cm,scale=.5,transform shape]
	\node [dot] (m2) at (0,0) {};
		\draw [->-] (1,0) to (m2);
\end{tikzpicture}, \begin{tikzpicture}[baseline=-0.1cm,scale=.5,transform shape]
	\node (mg2) [plus] at (0,0) {};
		\draw [->-] (1,0.5) to [out=180, in=45] (mg2);
		\draw [->-] (1,-0.5) to [out=180, in=-45] (mg2);
		\draw [->-] (mg2) to (-1, 0);
\end{tikzpicture}, \begin{tikzpicture}[baseline=-0.1cm,scale=.5,transform shape]
	\node (mg2) [circle, scale=0.5, draw] at (0,0) {$e$};
		\draw [->-] (mg2) to (-1,0);
\end{tikzpicture} \right) $ and $\left( \begin{tikzpicture}[baseline=-0.1cm,scale=.5,transform shape]
	\node [dot] (m2) at (0, 0) {};
		\draw [->-] (m2) [out=45, in=180] to (1,0.5);
		\draw [->-] (m2) [out=-45, in=180] to (1,-0.5);
		\draw [->-] (-1, 0) to (m2);
\end{tikzpicture} , \begin{tikzpicture}[baseline=-0.1cm,scale=.5,transform shape]
	\node [dot] (m2) at (0,0) {};
		\draw [->-] (-1,0) to (m2);
\end{tikzpicture}, \begin{tikzpicture}[baseline=-0.1cm,scale=.5,transform shape]
	\node (mg2) [plus] at (0,0) {};
		\draw [->-] (-1,0.5) to [out=0, in=135] (mg2);
		\draw [->-] (-1,-0.5) to [out=0, in=-135] (mg2);
		\draw [->-] (mg2) to (1, 0);
\end{tikzpicture}, \begin{tikzpicture}[baseline=-0.1cm,scale=.5,transform shape]
	\node (mg2) [circle, scale=0.5, draw] at (0,0) {$e$};
		\draw [->-] (mg2) to (1,0);
\end{tikzpicture} \right) $
satisfy the commutative bialgebra axioms in figure \ref{fig:commutative-bialgebra}, and they both interact as in figure \ref{fig:cup-equations} with the cup \begin{tikzpicture}[baseline=-1]
\draw [->-] (0,.6) to [out=0,in=90] (0.6,0) to [out=-90,in=0] (0,-0.6);
\draw [->-] (0,.3) to [out=0, in=90] (0.3,0) to [out=-90,in=0] (0,-0.3);
\end{tikzpicture}.\\
We explain in detail that the functor $F_\N$ preserves associativity of the black monoid, the rest of the equations are written with the same convention. We write on the left-most and right-most sides morphisms in $\Graph$ that, by associativity, they must be equal. In the centre, we write the morphisms in $\Game$ to which they are mapped (indicated with $\mapsto$) by the functor $F_\N$. These morphisms are equal in $\Game$ by associativity of the monoid operation $\oplus$ on $M$ and coassociativity of copying. Thus, we can say that  $F_\N$ preserves associativity of the black monoid.
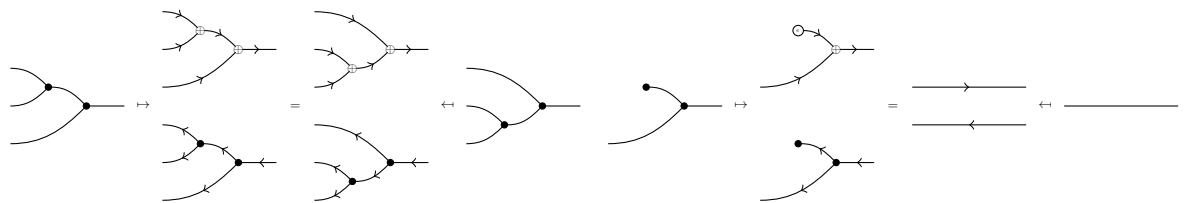
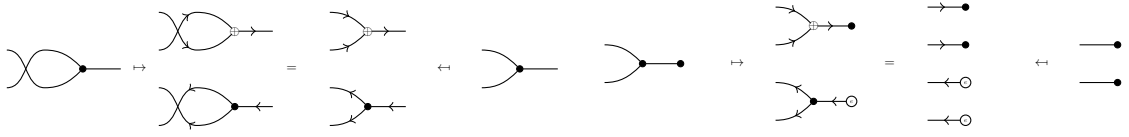
\monoidalfunctorproof
and similarly for their transposed versions.\\
\removed{\begin{center}
    \monoidalfunctorprooftransposed
  \end{center}}
\end{proof}
\removed{\begin{proof}[Proof of Corollary~\ref{th:closed_game_on_graph}]
$\Gamma = (k, \left[ ( \ ) \right], ( \ ), \text{!`}, !, \left[E \right] )$. Then, by Theorem~\ref{th:game_on_graph},
\begin{align*}
(\sigma, \sigma') \in \Bf_{F_{\mathcal{N}}(\Gamma)} \Leftrightarrow & \forall k' \ \sigma'_{k'} \in \operatorname*{argmax}_{s \in X} g( s, (E+E^T)\row{k'} f(\sigma \left[ k' \mapsto s \right] )) \\
\Leftrightarrow & \forall k' \ \sigma'_{k'} \in \operatorname*{argmax}_{s \in X} g (s, \bigoplus_{\substack{v_j v_{k'} \in E_\G \\ j \neq k'}} f (\sigma_j) \oplus  \bigoplus_{v_{k'} v_{k'} \in E_\G} f(s))
%\Leftrightarrow & \forall k' \ \sigma'_{k'} \in \operatorname*{argmax}_{s \in X} g( s, \bigoplus _{v_j v_k \in E_\G} ((1-\delta_{jk}) f(\sigma_j) \oplus \delta_{jk} f(s)))
\end{align*}
Thus, fix-points of $\Bf_{F_{\mathcal{N}}(\Gamma)}$ are Nash equilibria for $F_{\N}(\Gamma)$ and vice versa.
\begin{align*}
%(\sigma^*,\sigma^*) \in \Bf_{F_{\mathcal{N}}(\Gamma)} \Leftrightarrow & \forall k \ \forall s \in X \ g( \sigma^*, \bigoplus _{v_j v_k \in E_\Gamma} f(\sigma^*_j)) \geq g( s, \bigoplus _{v_j v_k \in E_\Gamma} ((1-\delta_{jk}) f(\sigma_j) \oplus \delta_{jk} f(s)))\\
(\sigma^*, \sigma^*) \in \Bf_{F_{\mathcal{N}}(\Gamma)} \Leftrightarrow & \forall k' \ \forall s \in X \ g( \sigma^*, \bigoplus_{v_j v_{k'} \in E_\Gamma} f(\sigma^*_j)) \geq g (s, \bigoplus_{\substack{v_j v_{k'} \in E_\Gamma \\ j \neq k'}} f (\sigma_j) \oplus \bigoplus_{v_{k'} v_{k'} \in E_\Gamma} f (s)) \\
\Leftrightarrow & \forall k' \ \forall s \in X \ u_{k'}(\Gamma,\sigma^*) \geq u_{k'}(\Gamma,\sigma^* \left[k' \mapsto s \right] ) \\
\Leftrightarrow & \sigma^* \text{ is a Nash equilibrium for }F_\N (\Gamma)
\end{align*}
\end{proof}}
\begin{proof}[Proof of Theorem~\ref{th:game_on_graph}]\label{proof:th_game_on_graph}
The proof proceeds by structural induction on $\Gamma$.\\
It is straightforward to check from the definition of $F_{\mathcal{N}}$ that the generators of $\Graph$ are sent to open games of the required form.\\
We need to check that composition is of the form as in the statement. We compute explicitly its play, coplay and best response functions.
\begin{align*}
&\Pf_{F_{\mathcal{N}}(\Gamma' \circ \Gamma)}((\vect{\sigma}, \vect{\sigma}'),\vect{x}) \removed{\\
&= \Pf_{F_{\mathcal{N}}(\Gamma') \cdot F_{\mathcal{N}}(\Gamma)}((\vect{\sigma}, \vect{\sigma}'),\vect{x}) \
&}=\Pf_{F_{\mathcal{N}}(\Gamma')}(\vect{\sigma}', B^T \vect{x} \oplus D^T f(\vect{\sigma})) \removed{\\
&= B'^T B^T \vect{x} \oplus B'^T D^T f(\vect{\sigma}) \oplus D'^T f(\vect{\sigma}') \
&}= (BB')^T \vect{x} \oplus \left( \begin{smallmatrix} DB' \\ D' \end{smallmatrix} \right)^T f( \left( \begin{smallmatrix} \vect{\sigma} \\ \vect{\sigma}' \end{smallmatrix} \right) )
\end{align*}
\begin{align*}
& \Cf_{F_{\mathcal{N}}(\Gamma' \circ \Gamma)}((\vect{\sigma}, \vect{\sigma}'),\vect{x},\vect{q}) \removed{\\
&= \Cf_{F_{\mathcal{N}}(\Gamma') \cdot F_{\mathcal{N}}(\Gamma)}((\vect{\sigma}, \vect{\sigma}'),\vect{x},\vect{q}) \\
&}= \Cf_{F_{\mathcal{N}}(\Gamma)}(\vect{\sigma}, \vect{x}, \Cf_{F_{\mathcal{N}}(\Gamma')}(\vect{\sigma}', \Pf_{F_{\mathcal{N}}(\Gamma)}(\vect{\sigma},\vect{x}),\vect{q}))\\
\removed{&= (A+A^T)\vect{x} \oplus B( (A'+A'^T)(B^T\vect{x} \oplus D^T f(\vect{\sigma})) \oplus B' \vect{q} \oplus C' f(\vect{\sigma}')) \oplus Cf(\vect{\sigma}) \\}
&= ((A + BA'B^T) + (A + BA'B^T)^T) \vect{x} \oplus BB'\vect{q} \oplus (C + B(A'+A'^T)D^T \vert BC') f( \left( \begin{smallmatrix} \vect{\sigma} \\ \vect{\sigma}' \end{smallmatrix} \right) )
\end{align*}
\begin{align*}
& ( \vect{\rho}, \vect{\rho}') \in \Bf_{F_{\mathcal{N}}(\Gamma' \circ \Gamma)}(\vect{x}, \cont) \\
\removed{\Leftrightarrow & ((\vect{\sigma}, \vect{\tau}), (\vect{\sigma}', \vect{\tau}')) \in \Bf_{F_{\mathcal{N}}(\Gamma') \circ F_{\mathcal{N}}(\Gamma)}(\vect{x}, \cont) \\}
\Leftrightarrow & (\vect{\sigma}, \vect{\sigma}') \in \Bf_{F_{\mathcal{N}}(\Gamma)}(\vect{x}, \cont \circ F_{\mathcal{N}}(\Gamma')_{\vect{\tau}}) \land (\vect{\tau}, \vect{\tau}') \in \Bf_{F_{\mathcal{N}}(\Gamma')}(F_{\mathcal{N}}(\Gamma)_{\vect{\sigma}} \circ \vect{x}, \cont) \\
\removed{\Leftrightarrow & \begin{multlined}[t]
	\forall k \ \sigma'_k \in \operatorname*{argmax}_{s \in X} g( s, (C^{T})\row{k} \vect{x} \oplus D\row{k} \Cf_{F_{\mathcal{N}}(\Gamma')}(\vect{\tau}, B^T\vect{x} \oplus D^T f(\vect{\sigma} \left[k \mapsto s \right]),\\
	\cont(\Pf_{F_{\mathcal{N}}(\Gamma')}(\vect{\tau}, B^T \vect{x} \oplus D^T f(\vect{\sigma} \left[ k \mapsto s \right])))) \oplus (E+E^T)\row{k} f(\vect{\sigma} \left[ k \mapsto s \right])) \\
	\land \forall k' \ \tau'_{k'} \in \operatorname*{argmax}_{s \in X} g( s, (C'^T)\row{k'} \Pf_{F_{\mathcal{N}}(\Gamma)}(\vect{\sigma}, \vect{x}) \oplus (D')\row{k'} \cont(B'^T \Pf_{F_{\mathcal{N}}(\Gamma)}(\vect{\sigma}, \vect{x})\\
	\oplus D'^T f(\vect{\tau} \left[k' \mapsto s \right])) \oplus (E+E'^T)\row{k'}f(\vect{\tau} \left[k' \mapsto s \right])) \\
	\end{multlined}\\
\Leftrightarrow & \begin{multlined}[t]
	\forall k \ \sigma'_k \in \operatorname*{argmax}_{s \in X} g( s, (C^T+D(A'+A'^T)B^T)\row{k} \vect{x}\\
	\oplus (D(A'+A'^T)D^T+E+E^T)\row{k}f(\vect{\sigma} \left[ k \mapsto s \right])\\
	\oplus (DC')\row{k}f(\vect{\tau}) \oplus (DB')\row{k} \cont ((BB')^T \vect{x} \oplus (DB')^T f(\vect{\sigma} \left[ k \mapsto s \right]) \oplus D'^T f(\vect{\tau}))) \\
	\land \forall k' \ \tau'_{k'} \in \operatorname*{argmax}_{s \in X} g( s, ((BC')^T)\row{k'} \vect{x} \oplus (E'+E'^T)\row{k'} f(\vect{\tau} \left[k' \mapsto s \right])\\
	\oplus ((DC')^T)\row{k'} f(\vect{\sigma}) \oplus (D')\row{k'} \cont ((BB')^T \vect{x} \oplus (DB')^T f(\vect{\sigma}) \oplus D'^T f(\vect{\tau} \left[k' \mapsto s \right])) \\
	\end{multlined} \\}
\Leftrightarrow & \begin{multlined}[t]
	\forall a=1,...,k+k' \\
	s \in \operatorname*{argmax}_{s \in X} g \Big( s, \begin{psmallmatrix} C^T + D(A'+A'^T)B^T \\ C'^T B^T \end{psmallmatrix} \row{a}\vect{x}\\
	\oplus \left( \begin{smallmatrix} DB' \\ D' \end{smallmatrix} \right)\row{a} \cont ( (BB')^T \vect{x} \oplus (B'^TD^T \vert D'^T) f(\vect{\rho} \left[ a \mapsto \rho'_a \right]))\\
	\oplus \left( \begin{smallmatrix} E+E^T+D(A'+A'^T)D^T & DC' \\ C'^T D^T & E'+E'^T \end{smallmatrix} \right) \row{a} f(\vect{\rho} \left[ a \mapsto \rho'_a \right])\Big)\\
	\end{multlined}\\
\end{align*}
Similarly, we show that monoidal product has the desired form.
\begin{align*}
&\Pf_{F_{\mathcal{N}}(\Gamma \otimes \Gamma')}((\vect{\sigma}, \vect{\sigma}'), (\vect{x},\vect{x}')) \removed{\\
&= \Pf_{F_{\mathcal{N}}(\Gamma) \otimes F_{\mathcal{N}}(\Gamma')}((\vect{\sigma}, \vect{\sigma}'), (\vect{x},\vect{x}')) \\
&= (B^T \vect{x} \oplus D^T f(\vect{\sigma}), B'^T\vect{x}' \oplus D'^T f(\vect{\sigma}')) \\ 
&}= \left( \begin{smallmatrix} B & \mathbb{0} \\ \mathbb{0} & B' \end{smallmatrix} \right) ^T \left( \begin{smallmatrix} \vect{x} \\ \vect{x}' \end{smallmatrix} \right) \oplus \left( \begin{smallmatrix} D & \mathbb{0} \\ \mathbb{0} & D' \end{smallmatrix} \right) ^T f( \left( \begin{smallmatrix} \vect{\sigma} \\ \vect{\sigma}' \end{smallmatrix} \right))
\end{align*}
\begin{align*}
&\Cf_{F_{\mathcal{N}}(\Gamma \otimes \Gamma')}((\vect{\sigma}, \vect{\sigma}'), (\vect{x},\vect{x}'), (\vect{r},\vect{r}')) \\
\removed{&= \Cf_{F_{\mathcal{N}}(\Gamma) \otimes F_{\mathcal{N}}(\Gamma')}((\vect{\sigma}, \vect{\sigma}'), (\vect{x},\vect{x}'), (\vect{r},\vect{r}')) \\
&= ((A+A^T)\vect{x} \oplus B\vect{r} \oplus Cf(\vect{\sigma}), (A'+A'^T)\vect{x}' \oplus B'\vect{r}' \oplus C'f(\vect{\sigma}')) \\ }
&= \left( \left( \begin{smallmatrix} A & \mathbb{0} \\ \mathbb{0} & A' \end{smallmatrix} \right) + \left( \begin{smallmatrix} A & \mathbb{0} \\ \mathbb{0} & A' \end{smallmatrix} \right)^T \right) \left( \begin{smallmatrix} \vect{x} \\ \vect{x}' \end{smallmatrix} \right) \oplus \left( \begin{smallmatrix} B & \mathbb{0} \\ \mathbb{0} & B' \end{smallmatrix} \right) \left( \begin{smallmatrix} \vect{r} \\ \vect{r}' \end{smallmatrix} \right) \oplus \left( \begin{smallmatrix} B & \mathbb{0} \\ \mathbb{0} & B' \end{smallmatrix} \right)  f( \left( \begin{smallmatrix} \vect{\sigma} \\ \vect{\sigma}' \end{smallmatrix} \right))
\end{align*}
\begin{align*}
& ( \vect{\rho}, \vect{\rho}') \in \Bf_{F_{\mathcal{N}}(\Gamma' \otimes \Gamma)}((\vect{x},\vect{x}'), \langle \cont, \cont' \rangle) \\
\removed{\Leftrightarrow & ((\vect{\sigma}, \vect{\tau}), (\vect{\sigma}', \vect{\tau}')) \in \Bf_{F_{\mathcal{N}}(\Gamma') \otimes F_{\mathcal{N}}(\Gamma)}((\vect{x},\vect{x}'), \langle \cont, \cont' \rangle) \\
\Leftrightarrow & (\vect{\sigma}, \vect{\sigma}') \in \Bf_{F_{\mathcal{N}}(\Gamma)}(\vect{x}, \cont(-,\Pf_{F_{\mathcal{N}}(\Gamma')}(\vect{\tau},\vect{x}'))) \land (\vect{\tau}, \vect{\tau}') \in \Bf_{F_{\mathcal{N}}(\Gamma')}(\vect{x}', \cont'(\Pf_{F_{\mathcal{N}}(\Gamma)}(\vect{\sigma},\vect{x}),-)) 	\\
\Leftrightarrow & \begin{multlined}[t]
	\forall k \ \sigma'_k \in \operatorname*{argmax}_{s \in X} g( s, (C^T)\row{k}\vect{x} \oplus D\row{k} \cont(B^T \vect{x} \oplus D^T  f(\vect{\sigma} \left[ k \mapsto s \right]), B'^T\vect{x}' \oplus D'^T f(\vect{\tau})) \\
	\oplus (E+E^T)\row{k} f(\vect{\sigma} \left[ k \mapsto s \right])) \\
	\land \forall k' \ \tau'_{k'} \in \operatorname*{argmax}_{s \in X} g( s, (C'^T)\row{k'}\vect{x}' \oplus (D')\row{k'} \cont'(B^T\vect{x} \oplus D^T f(\vect{\sigma}), B'^T \vect{x}' \oplus D'^T f(\vect{\tau} \left[ k' \mapsto s \right]))\\
	\oplus (E'+E'^T)\row{k'}f(\vect{\tau} \left[ k' \mapsto s \right]))\\
	\end{multlined} \\}
\Leftrightarrow &
	\forall a=\,...,k+k' \
	\rho'_a \in \operatorname*{argmax}_{s \in X} g\Big( s, \left(\left( \begin{smallmatrix} C & \mathbb{0} \\ \mathbb{0} & C' \end{smallmatrix} \right)^T\right)\row{a} \left( \begin{smallmatrix} \vect{x} \\ \vect{x}' \end{smallmatrix} \right)
  \oplus \left(\left( \begin{smallmatrix} D & \mathbb{0} \\ \mathbb{0} & D' \end{smallmatrix} \right)^T\right)\row{a} \langle \cont, \cont' \rangle (\left( \begin{smallmatrix} B & \mathbb{0} \\ \mathbb{0} & B' \end{smallmatrix} \right)^T \left( \begin{smallmatrix} \vect{x} \\ \vect{x}' \end{smallmatrix} \right) \\
  &\oplus \left( \begin{smallmatrix} D & \mathbb{0} \\ \mathbb{0} & D' \end{smallmatrix} \right)^T f(\vect{\rho} \left[ a \mapsto s \right]))
	\oplus \left( \left( \begin{smallmatrix} E & \mathbb{0} \\ \mathbb{0} & E' \end{smallmatrix} \right) + \left( \begin{smallmatrix} E & \mathbb{0} \\ \mathbb{0} & E' \end{smallmatrix} \right)^T \right) \row{a} f(\vect{\rho} \left[ a \mapsto s \right])\Big)
\end{align*}
\end{proof}

\end{document}